\documentclass[manuscript]{aastex}

\newcommand{\mua}{$\mu_\alpha \cos \delta$}
\newcommand{\mud}{$\mu_\delta$}
\newcommand{\qso}{\object{QSO J0240-3434B}}
\newcommand{\masy}{\mbox{mas y$^{-1}$}}
\newcommand{\kms}{\mbox{km s$^{-1}$}}

\slugcomment{Accepted by PASP.}

\shorttitle{Fornax in motion}
\shortauthors{M\'endez et al.}

\begin{document}

\title{Proper motions of Local Group dwarf spheroidal galaxies I:
  First ground-based results for Fornax\footnote{Based on observations
    made with ESO Telescopes at the La Silla Observatory under
    programmes ID $<060.A-9701>$, $<065.N-0472>$, $<066.B.0276>$,
    $<074.B-0456>$, $<076.B-0528>$, $<079.B-0379>$, $<080.B-0144>$,
    $<081.B-0534>$, and $<082.B-0034>$.}}

\author{Ren\'e. A. M\'endez and Edgardo Costa}
\affil{Departamento de Astronom\'ia, Universidad de Chile, Casilla
  36-D, Santiago, Chile}
\email{rmendez@u.uchile.cl, costa@das.uchile.cl}

\author{Mario H. Pedreros}
\affil{Departamento de F\'isica, Universidad de Tarapac\'a, Casilla 7-D, Arica, Chile}
\email{mpedrero@uta.cl}

\author{Maximiliano Moyano}
\affil{Max-Planck-Institut f\"ur Astronomie, K\"onigstuhl 17, D-69117,
  Heidelberg, Germany}
\email{moyano@mpia-hd.mpg.de}

\author{Martin Altmann}
\affil{Zentrum f\"ur Astronomie,
  M\"onchhofstr. 12-14, 69120 Heidelberg, Germany}
\email{maltmann@ari.uni-heidelberg.de}

\and

\author{Carme Gallart}
\affil{Instituto de Astrof\'isica de Canarias, Tenerife 38200, Islas Canarias, Spain}
\email{carme@iac.es}

\begin{abstract}

In this paper we present in detail the methodology and the first
results of a ground-based program to determine the absolute proper
motion of the Fornax dwarf spheroidal galaxy.


The proper motion was determined using bona-fide Fornax star members
measured with respect to a fiducial at-rest background
spectroscopically confirmed Quasar, \qso. Our homogeneous
measurements, based on this one Quasar gives a value of (\mua,\mud)$ =
(0.64 \pm 0.08, -0.01 \pm 0.11)$~\masy. There are only two other
(astrometric) determinations for the transverse motion of Fornax: one
based on a combination of plates and HST data, and another (of higher
internal precision) based on HST data. We show that our proper motion
errors are similar to those derived from HST measurements on
individual QSOs. We provide evidence that, as far as we can determine
it, our motion is not affected by magnitude, color, or other potential
systematic effects. Last epoch measurements and reductions are
underway for other four Quasar fields of this galaxy, which, when
combined, should yield proper motions with a weighted mean error of
$\sim50\,\mu$as~y$^{-1}$, allowing us to place important constraints
on the orbit of Fornax.

\end{abstract}


\section{Introduction}

The proper motions (PMs) of the satellites of the Milky Way (MW), when
combined with existing radial velocities, allow us to determine the
space velocity vectors of these satellites of our galaxy, which in
turn place important constraints on their orbits (see, e.g.,
\citet{bes07}, \citet{pia07}). This knowledge is crucial to determine
if these galaxies are gravitationally bound to the Galaxy, and to our
understanding of the evolution and origin of its satellite system
(\citet{byr94}). The PMs of the satellites of the MW are necessary to
understand: a) the origin of the MW satellite system and its
relationship with the formation and evolution of the galactic halo
(\citet{din00}, \citet{pal02}, \citet{car09}), b) the nature and
origin of the streams that seem to align different subgroups of these
galaxies (\citet{lyn82}, \citet{lyn95}, \citet{pia05}), and c) the
role of tidal interactions in the evolution and star formation history
of low mass galaxies (\citet{zar04}, \citet{pia05}, \citet{noe09},
\citet{may10}). A comprehensive study of the MW's satellite system can
lead to a greater general understanding of galaxy evolution and the
physical processes governing star formation in galaxies.


From another perspective, reliable space motions of Local Group
galaxies is a key ingredient to populate the phase-space components
for flow models which predict the dynamical evolution of the local
universe. Indeed, current center-of-mass locations and motions from
the present distribution of galaxies can be used as boundary
conditions to ``trace them back in time'' \citep{pee89,pee94} and to
test the paradigm that galaxy clusters grew by gravitational
instabilities from an originally smoother medium, with small random
motions. These earlier simulations have not been repeated too often
due to the growing realization that more complex effects such as
tides, satellite interactions, accretion and mergers, influencing the
growth of galaxies with time (specially at epochs earlier than 7-8~Gyr
from now), can greatly complicate this approach. Nevertheless, the
motions of local group galaxies can be used as boundary conditions in
a first aproximation to carry out the above analysis. This goal has
become one of the important drivers behind future astrometric space
missions, such as SIM (\citet{unw08}, specially their Figure~13),
which expect to measure PM for Local Group galaxies with a precision
ten times better than what can be achieved with present techniques.

With the above motivations in mind, in the year 2000 we started a
ground-based program aimed at determining,
the absolute PM of three southern dwarf Spheroidal (dSph) galaxies,
Carina, Fornax, and Sculptor, with respect to known background Quasars
(QSOs) that can be used as inertial reference points. Three epochs,
over a period of eight years were obtained using a single
telescope+detector set up: Ours is then the first entirely optical
CCD/ground-based proper motion study of an external galaxy other than
the Magellanic Clouds.

In this paper we report on the first results from this program, based
on one QSO field in Fornax, for which we have data of good enough
quality to allow us to asses the expected precision of our
measurements, and to describe our methodology in detail. Last epoch
measurements and reductions are underway for other four Quasar fields
of Fornax, as well as for a similar number of QSO fields in Carina and
Sculptor. The results for these will be presented in forthcoming
papers.

In section~\ref{obma} we describe our observational material and data
acquisition strategies, in section~\ref{rest} we describe our
methodology for deriving the PMs, in section~\ref{anal} we present the
analysis for our QSO field, and in section~\ref{comp} we present our
main conclusions as well as a comparison to other results.

\newpage
\section{Observational material} \label{obma}

All our observations were carried out with the ``Super Seeing
Imager'', SuSI2, attached to one of the Nasmyth focii of the ESO 3.5 m
NTT telescope at La Silla Observatory. The overall characteristics of
the detector, a mosaic of two EEV 44-82, 2k$\times$4k, 15~$\mu$m
pixel, thinned, anti-reflection coated chips covering a field-of-view
(FOV) of 5.5$\times$5.5 arcmin$^2$, is fully described in
\citet{dod98a} and \citet{dod98b}. The measured charge transfer
efficiency, CTE (serial and parallel readouts) is better than 1 part
in $10^6$, while both chips have been found to be linear within
$\pm$0.15\% over the full range 0-60,000~ADUs. All our measurements
are based on objects within this linearity range. On the other hand,
CTE is considered to be have a negligible impact on our astrometric
measurements, as the sky background (unlike the case of the
space-based, e.g., HST, data) is quite large, typically between 300
and 900 ADUs (in comparison to the few counts on the HST cameras):
Higher background signals fills-in charge traps on the detector,
leaving no empty traps to be filled by signal from a source (for
details see, e.g., \citet{bri06}, specially his Figure~10).


In the un-binned mode, in which all our data was acquired, we have
adopted for the plate scale the value given in the SuSI2 manual 
of $0.0805 \pm 0.0002$~arcsec~pix$^{-1}$. We have verified the
accuracy of this value, following the procedure described in
\citet{cos09}, through the use of the IRAF routines {\tt ccxymatch,
  ccmap} and {\tt cctran} on an astrometric field in the periphery of
the $\Omega$Cen cluster \citep{van00}, with plenty of stars in the
FOV. All the astrometric and refraction-series frames (see
section~\ref{anal}) were acquired through a ``Bessel R \#813'' filter,
whereas blue frames needed for constructing color-magnitude diagrams
(CMDs) used the ``Bessel B \#811'' filter. Both filters are part of
the standard set for the instrument. 

The QSO field analyzed here is that of the double QSO~J0240-3434,
located at $(RA,DEC)=(02:40:07.7,-34:34:20)$ (position for the A
component, although for reasons that are explained later one we used
only the B-component for our PM, see
section~\ref{centroids}). Exposure times for the science frames varied
between 250~s and 900~s (see Table~\ref{frames}). At the start of our
program we adjusted the exposure time on the basis of the ambient
seeing, later it was decided to use a fixed exposure (900~s) in all
cases.

Because our measurements involve the displacement of Fornax stars
relative to a fixed point (the QSO in the FOV), we used only one of
the two chips of the array, Chip \#46, which has a slightly better
read-out noise (4.6~e$^-$/pix, with a gain of 2.26~e$^-$/ADU), and
better overall cosmetics than its twin. Given the alt-az nature of the
telescope, and the Nasmyth location of the instrument, a
rotator-adaptor allows one to select the orientation of the detector
on the sky, and keeps it fixed during the observations. All our
observations were acquired with a nominal rotation angle of zero
degrees: In this orientation North increases in the Y-pixel direction,
while East increases in the X-pixel direction (for more details see
section \ref{building}).

For reasons that will be become apparent later on, all observations
were acquired by placing the QSO at a nominal pixel position of
$(x,y)=(3\,000,\,2\,100)$ near the middle of chip \#46. Given the
extraordinary pointing accuracy of the telescope, this meant that all
our frames are centered at this position $\pm$~few pix (few=$ \le
7$~pix~$\sim0.6$~arcsec). This procedure was greatly simplified by the
use of the so-called ``Observing
Blocks''\footnote{http:$/$$/$www.eso.org$/$sci$/$facilities$/$lasilla$/$instruments$/$susi$/$docs$/$SUSItemplates.html}
that determine the way in which a set of observations with a given
telescope+instrument is to be carried out, a common feature of the
data acquisition software at all ESO facilities (other relevant
factors such as hour angle, airmass, and the range of acceptable
seeing and sky transparency conditions were however decided by the
observer).

For calibration purposes at least five dome illumination flats in the
B and R filters were secured every night. Some sky illumination flats
were also acquired to check our basic calibrations (see below),
pointing the telescope one hour to the East/West (sunset/dawn) of the
meridian, and at a declination equal to the latitude of the
observatory, always with the telescope tracking, and applying small
offsets in between exposures to facilitate the elimination of any
possible stellar images during the combination of the images. Also, at
least twenty zero exposure time bias frames were obtained every
night. Given the negligible dark current of the instrument, no dark
frames were acquired. The mosaic data was split into the two
individual chips, and only data from chip \#46 was used
subsequently. All frames (from now on single-chip \#46 exclusively)
were calibrated, on a night-by-night basis, using standard IRAF
routines within the {\tt noao.imred.ccdred} package for combining the
zero \& flats frames, and applying them to the science images. In
particular, zero and illumination flat frames were median-combined
with an iterative "avsigclip" algorithm with an upper \& lower sigma
clipping of 2.5. For combination, the flats were scaled using their
respective median values computed from a region free from cosmetic
defects in the middle of the chip, encompassing
1300$\times$3500~pix$^2$ (no scaling was used for the zero exposure
frames, as their level remained constant, save by Poisson noise
fluctuations).

An overscan region, resulting from the average of 38 non-exposed
columns to the right of chip \#46, was fitted with three pieces of a
third degree spline function, which was subsequently subtracted from
all frames. This function provided a smooth and well sampled fit to
the observed overscan all the way to the usable edges of the chip:
Starting from a full exposed CCD area 2048$\times$4096~pix$^2$ (each
chip has in total 2144$\times$4096~pix$^2$), the final trimmed images
have 2026$\times$4086~pix$^2$, where rapidly variable illumination
and/or CCD boundary defects were excluded altogether.

Whenever we had a reasonable number of well-exposed sky flats (at
least three), the science data was reduced with sky and dome flats
independently to asses any large remaining gradients that may be
induced by a non-uniform illumination from the dome flat screen. These
tests indicated that, after flat-fielding, there remained
low-frequency trends across the chip using {\it either} dome or sky
flats, and that sky flats were not always necessarily better than dome
flats, perhaps as a consequence of scattered light coming into the
detector from the telescope structure (through its wind vents) at
sunset/dawn. Larger trend were noticed in our earlier epoch data. Over
the years a number of improvements were made to the telescope
baffling\footnote{See, e.g.,
  http:$/$$/$www.eso.org$/$sci$/$facilities$/$lasilla$/$sciops$/$doc$/$LSO\-TRE\-ESO\_40200-1061\_susiBaffle$/$LSO-TRE-ESO\_40200-1061\_susiBaffle.html},
which alleviated these problems in the later epoch frames, as indeed
observed in our data. In any case, the maximum background gradient
found in these comparisons was 10\% accross the entire chip (although
typically it was $\sim$5\% or smaller), and we did not find any large
gradients on scales smaller than a few hundred pixels. Therefore,
locally, at any location on the chip, the sky could be considered
essentially constant. Since our primary concern is astrometry, and not
photometry, we opted for reducing all our data (i.e., B and R-band
frames) consistently using dome flats only.

After the basic calibrations described above were performed for the
data from all epochs, each frame was used independently to determine
$(x,y)$ positions and photometry for a carefully selected subset of
objects in the FOV, thus starting the astrometric reduction
steps described in detail in the following section.

\newpage
\section{Astrometric reduction steps} \label{rest}

The basic reduction steps are similar to those described in
\citet{cos09} in the context of the so-called ``QSO method'' in which
the relative displacement of a fiducial at-rest point (the QSO) is
measured with respect to a set of bona-fide Fornax stars, registered
to a common reference system.
In what follows we give further details on either departures from
the basic astrometric steps described by \citet{cos09}, or provide
more detailed explanations whenever required.

After grouping all (astrometric) data for a given QSO field, we
created a catalog of all observations with basic bookkeeping
information such as date of observation, exposure time, hour angle,
airmass, etc., of the images. We then examined all images to determine
their overall quality, including average seeing ($FWHM$), sky
background ($<sky>$), and sky noise ($\sigma_{\mbox{sky}}$),
information which was cross-correlated with the night sky conditions
as recorded on the various observatory logs\footnote{See, e.g.,
  http:$/$$/$www.eso.org$/$gen-fac$/$pubs$/$astclim$/$lasilla$/$index.html}
and by the DIMM seeing associated to each image as recorded on the ESO
Science
Archive\footnote{http:$/$$/$archive.eso.org$/$asm$/$ambient-server}. This
first quality control allowed us to discard unusable images (e.g.,
trailed or badly elongated images, very low counts due to variable sky
transparency, and ``bad'' ($> 1.2$~arcsec seeing)).
In Table~\ref{frames} we summarize the observational material used
throughout this work. The meaning of the various entries will be
explained in the following subsections.

\subsection{Building the initial astrometric reference system \& deriving pixel coordinates} \label{building}

From the basic data set obtained as above, we selected a set of
good-seeing ($FWHM < 0.7$~arc-sec), near meridian, consecutive frames
to create the so-called ``Standard Frame of Reference'' (SFR, which
will be used later on in the astrometric registration process), and
selected one of them as a ``master'' frame (see
Table~\ref{frames}). From this master frame we created a list of
``good'' reference stellar images to define our initial ``local
reference system'' (``LRS'' from now on). To create the LRS we used
the DAOPHOT tasks implented in the IRAF package under {\tt
  noao.digiphot.daophot}. We start by running the {\tt daofind} task
with a small threshold of 4$\times \sigma_{\mbox{sky}}$,
to have as many detections as possible (it is important to start
with a low threshold, since it allows us to use these hits to
eliminate potential nearby companions to much brighter good reference
stars, see below). Proper care was taken to edit relevant image
parameters such as the mean FWHM of the stellar images,
$\sigma_{\mbox{sky}}$, the minimum and maximum good data value (taken
as $<sky>-10\times \sigma_{\mbox{sky}}$ and 60,000~ADUs
respectively). The output of {\tt daofind} had, at this stage 8,235
hits, in total.

We note that the orientation of the master frame with respect to the
Celestial RA, DEC coordinate system defines also the orientation of
our PMs with respect to \mua\ and \mud\ since both the individual SFR
frames and all of the other astrometric frames are eventually
transformed into the master frame (see
section~\ref{method}). Therefore, it is important to ascertain any
local ``rotation'' of our $(x,y)$ coordinates with respect to
$(\alpha,\delta)$. This was done following the same procedure
described in \citet{cos09}: We found a negligible rotation of our
master frame ($359.545 \pm 0.018$~deg), and therefore no further
correction (which go as the {\it cosine} and {\it sine} of the above
angle) was applied to our PMs. Since the NTT is an Alt-AZ telescope,
and the instrument is mounted on one of the Nasmyth focii, we were
concerned that the adaptor-rotator could have introduced significant
rotations between succesive frames due to mechanical innacuracies in
the positioning of the rotator. When explicitely solving for this
remaining rotator-induced rotation we found that they were all very
small (the largest measured rotation was 0.017~deg, although it was
typically 0.0001~deg). In any case, we note that our registration
procedure (in particular the first-order cross terms) would fully
account for this effect when comparing the SFR to all other individual
astrometric frames.

Photometry with the task {\tt phot} was then performed using the input
list described above on this particular master frame, and using a
rather large aperture of 15~pix (1.2~arcsec). To determine the local
sky, an annulus of width 5~pix and inner radius of 20~pix was
adopted. An iterative sigma-clipping algorithm with upper \& lower
rejection limit of 2.5, as well as an upper \& lower clipping factor
of 10\% percent, was adopted to robustly determine the mean local
value of the sky around each object, as measured from the modal value
(recommended for crowded fields). The output of the {\tt phot} task
was then used to make a preliminary semi-automatic ``cleaning'' of the
sample by eliminating objects, using the {\tt pexamine} tool (given
the larger number of hits, at this stage we departed from the
\citet{cos09} procedure of examining the possible LRS stars by eye,
which we did only at a later time). Objects were deleted either
because they were too close to the frame boundaries (and thus could
become missing in slightly offset frames from other epochs), or
because they were at or near a set of two bad columns in the upper
part of the chip. Additionally, image shape or photometric parameters
reported by the {\tt daofind} and {\tt phot} tasks, such as
``sharpness'', ``roundness'', and magnitude error were used as a
preliminary guide to select a set of well measured star images on the
master frame. Spurious detections around bright stars, as well as
multiple hits on extended objects were also eliminated based on an eye
inspection of the {\tt daofind} hits on the master frame. At this
stage the number of detections dropped to some 952 objects. We then
ran a code to eliminate, out of this latest sample, any object that
had a companion in the original {\tt daofind} output, closer than a
certain radius, in this case, 20~pix, rendering the number of objects
down to 422. The choice of this radius is somewhat arbitrary and is a
compromise between a large enough number of LRS stars (i.e., choosing
a small radius), and a set of well isolated stellar images for an
astrometrically stable LRS (free from the influence of
nearby/partially blended images, i.e., large radius), which is a
function of the stellar density of the particular image. We note for
example that when increasing this radius to 30~pix the number of
objects dropped to 79 stars, which we considered insufficient given
the further cleansing applied to the LRS (see below).

A small complication arose in the selection of the LRS stars since the
Fornax Globular Cluster ``4''\footnote{according to the SIMBAD
  database} at RA=02:40:07.70, DEC=-34:32:09.8, first described by
\citet{hod61} was present in the images in the North section of the
chip. The cluster was not well resolved in the images, leading to a
large number of stars with nearby companions in a crowded field. To
avoid potential astrometric problems with these stars, a reasonable
region around the cluster was removed. In the end, this step did not
have an important impact on the x-y distribution of our reference
stars (see Figure~\ref{xylrs}).

With this list we then interactively ran the tasks {\tt pstselect} and
{\tt psf} and selected a subset of high S/N stars well distributed
over the entire chip, to build a point spread function (PSF) for the
master image. We let DAOPHOT decide what psf function to adopt, based
on the best $\chi^2$ of the fit, as reported by IRAF, on account of
the differing seeing/distortions in different frames. Given the large
number of stars available, we also let the PSF to vary accross the
chip (fitting the lookup parameters to each individual image), by
adopting {\tt varorder=3} within the {\tt daopars} parameter set which
is used to drive the {\tt psf} task. One advantage of running these
taks interactively is that stars with previously unseen companions or
with cosmic rays and/or hot pixels can be excluded in the construction
of the psf (stars with previously undetected close companions or
affected by hot pixels were also excluded from the LRS list from which
the psf list was built, but not the ones affected by cosmic-rays,
which are image-dependent). As an example, for the master frame, out
of 200 initially selected PSF stars, only 170 survived this
interactive cleaning, and were used to create the PSF.

Once a PSF for the master frame was created, the iterative PSF-fitting
DAOPHOT task {\tt peak} was used to compute final $(x,y)$ coordinates
for all the LRS stars. Two further cleaning steps of the initial LRS
were performed at this stage. First, from the {\tt peak} output
(which, on account of using a PSF built from the image itself,
provides a much refined measurement of image parameters than {\tt
  daofind} and {\tt phot} do), objects with large magnitude errors
($\sigma_{\mbox{mag}} > 0.1$~mag), and/or with an unusually large
number of iterations (in comparison with the bulk of objects) to
converge to the solution were deleted. Finally, this latter list was
applied to the worst FWHM astrometric image of this QSO field to
create a PSF and to run {\tt peak}, again deleting objects that, on
this relatively lesser quality image, had parameters similar to those
used to eliminate objects in the (good quality) master frame. At this
stage we had built a sample of 260 well measured stars over the entire
FOV of the chip which became our initial LRS for the entire set of
frames. The approximate magnitude and color range for the LRS is $17.0
\le R \le 21.2$ and $0.0 \le B-R \le 2.0$ (see Section~\ref{anal}).

Once the initial LRS was built, we succesively created PSFs for all
frames, always in interactive mode to eliminate stars affected by
cosmic-rays (therefore number of PSF stars was variable, between a
minimum of 128 and a maximum of 178 stars), and used them to compute
$(x,y)$ coordinates for all the LRS stars and the QSO on each frame
using the task {\tt peak}. This was perhaps the most time-consuming
part of the whole process, 
given its interactive nature, and that for each frame the {\tt psf}
task had to compute the goodness-of-fit for the six functions
available within DAOPHOT: Gaussian, two Moffat functions, two Penny
functions, and one Lorentzian). For each frame we kept the radius of
PSF model fixed at 14~pix, whereas the fitting radius was adjusted to
$1.2 \times FWHM$ of the corresponding frame, with a maximum value
equal to the adopted PSF model radius. The ouput of the {\tt peak}
routine are the raw pixel coordinates of the LRS stars on each frame,
which provide the basic ingredient for the subsequent astrometric
reduction steps, described in the following subsections.

\subsection{Precision of the pixel coordinates} \label{centroids}

Numerous tests were performed to determine the best precision
attainable for our raw pixel coordinates as a function of the various
PSF fitting parameters. First, given the degree of arbitrariness with
which one includes or excludes PSF stars on each frame, we tested the
sensitivity of our derived pixel coordinates to the exact nature of
the stars chosen to build the PSF (ideally, one would like to have the
same basic set of PSF stars for all frames, but the appearance of
comsic rays - different on each image - on otherwise good PSF stars
dot not allow this). In general, for all our frames, we started with
the maximum number of stars (200) allowed by the {\tt psf} task, and
interactively deleted bad PSF stars, as explained before, but without
regard of their S/N. The main motivation for this, besides the large
number of suitable stars available, is to have a well sampled PSF over
the entire FOV. Using one of the best FWHM consecutive frames
indicated as ``CTB-1'' in Table~\ref{frames}, we computed a PSF (from
135 stars) and $(x,y)$ {\tt peak}-based coordinates for a total sample
of 1,200 stars, covering a wide magnitude range, and going actually
$\sim$2~mag fainter than our LRS (equivalent to $\sim$1~mag brighter
than the 4$\sigma_{\mbox{sky}}$ {\tt daofind} detection limit for this
frame), using our standard procedure: For these frames, this meant
that the faintest PSF stars had a peak value above sky of $\sim900$
counts (equivalent to an overall S/N over the whole image of $\sim200$
on an aperture radius of $2 \times FWHM$). We then computed another
PSF but this time based only on stars with peak counts above
$\sim2000$ (in total the brightest 49 stars of the sample used to
created the previous PSF), following the arguments proposed by
\citet{and06} (see, in particular, their Section~3.5), and used this
PSF to get $(x,y)$ coordinates for the same 1,200 stars. A comparion
of the coordinates between the two sets indicate minuscule offsets of
$<\Delta X> = (0.03 \pm 0.84) \times 10^{-3}$~pix, and $<\Delta Y> =
(-0.03 \pm 1.0) \times 10^{-3}$~pix. The offsets do not show any trend
{\it vs.}  position, magnitude, or color of the stars. As we shall see
later on, these offsets are a factor of 600 smaller than our (single
measurement) centering accuracy when comparing consecutive frames (in
time), and we therefore conclude that our choice of PSF stars is not
biasing our pixel coordinates in any measurable way.

\begin{figure}
\epsscale{.80}
\plotone{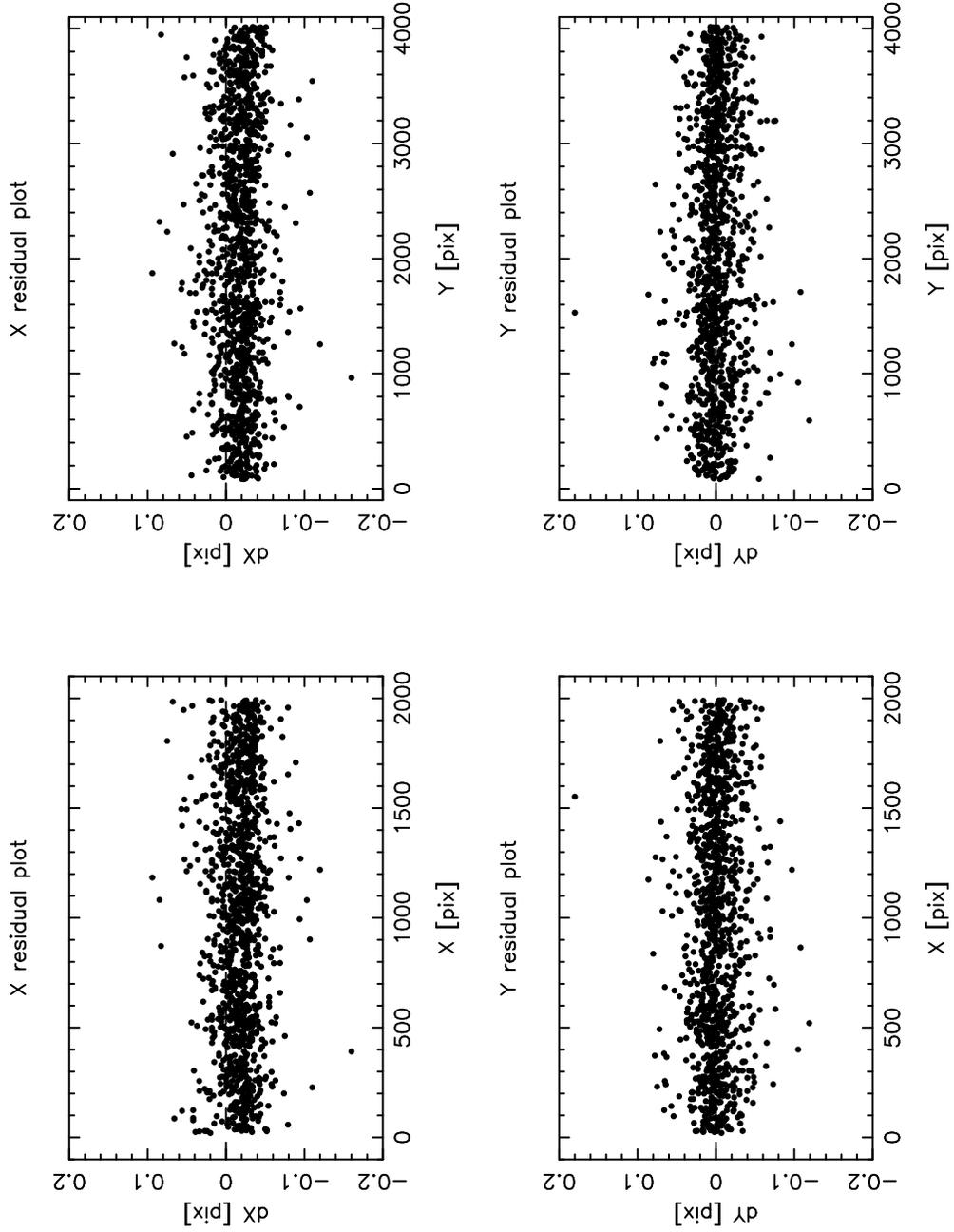}
\caption{Coordinate shifts for the same frame as measured by using two
  extremely different PSF functional forms. No positional residual
  trends can be seen. \label{psftrends1}}
\end{figure}

\begin{figure}
\epsscale{.80}
\plotone{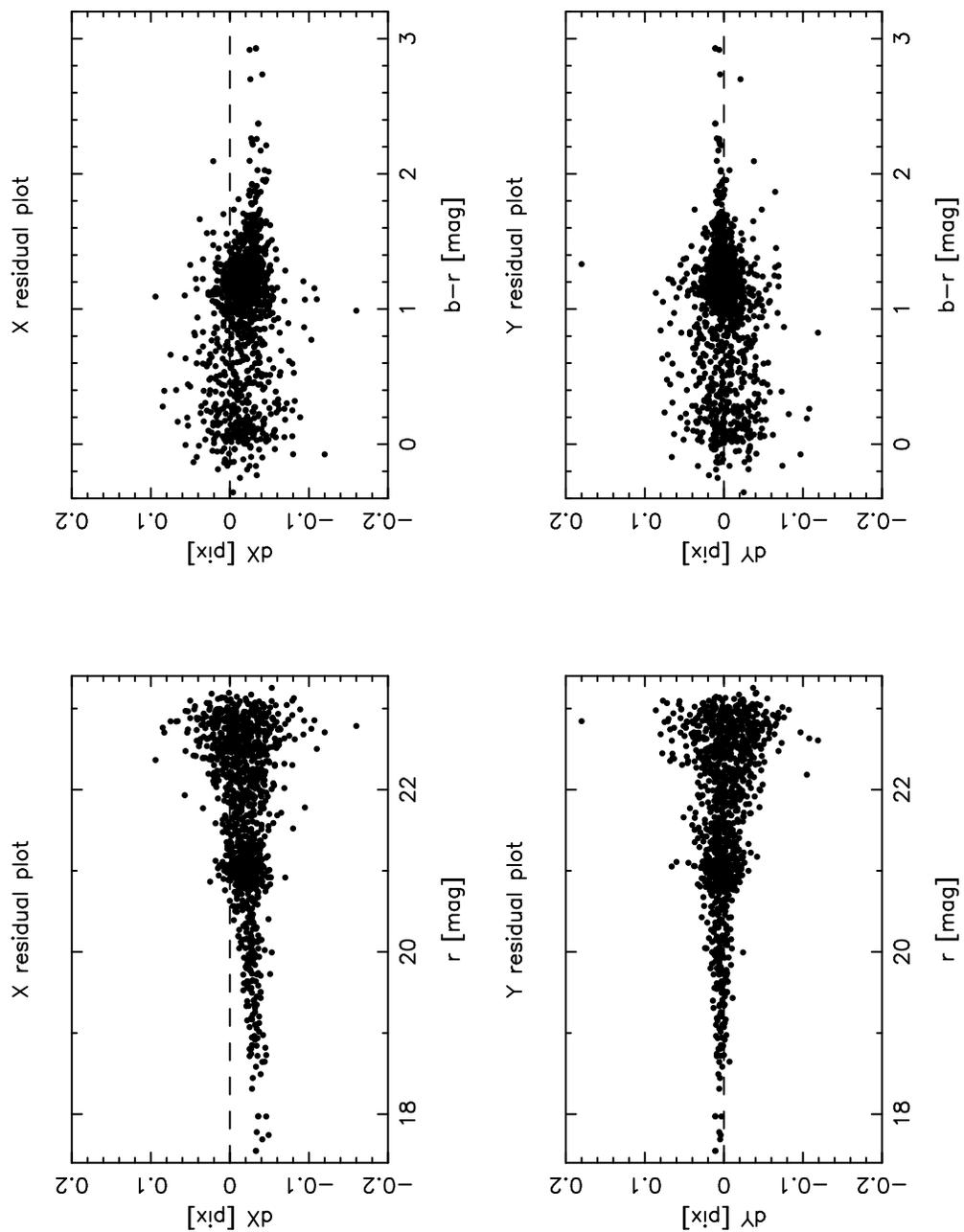}
\caption{Same as Figure~\ref{psftrends1} but as a function of
  instrumental r-band magnitude and b-r color. A clear trend in the
  X-coordinate is found as a function of magnitude, and a trend as a
  function of color is suggested. \label{psftrends2}}
\end{figure}

In our standard procedure we allow the software to decide (based on a
$\chi^2$ test) what functional form for the PSF to adopt (in the case
of frame CTB-1 this was a ``penny2'', with 6 lookup tables), and we
decided to also estimate the impact of this choice in our
coordinates. For this purpose we fixed the PSF to the various
functional forms allowed by the {\tt daophot} task, but always using
the same PSF stars and PSF parameters. The worst-case scenario was
found when comparing the $(x,y)$ coordinates to those derived from a
``gaussian'' PSF with no spatial variation accross the chip. In this
case, while no obvious positional trends are found
(Figure~\ref{psftrends1}), a magnitude dependent systematic trend
(beyond the obvious increase in the scatter at fainter magnitudes) is
clearly visible, (Figure~\ref{psftrends2}) amounting to $(4.67 \pm
0.61) \times 10^{-3}$~pix~mag$^{-1}$. This trend disappears {\it for
  any other PSF functional form}. Even, with the next degree of
freedom, i.e., a simple Gaussian PSF with only one lookup table (and
no positional variation of the PSF accros the chip), we find a slope
(in the X-coordinate) of $(-2.9 \pm 2.4) \times
10^{-4}$~pix~mag$^{-1}$, i.e., consistent with zero. We conclude that
the trend is due to the inadequacy of a simple analytical PSF to model
the actual PSF in the image, but that our exact choice of the PSF is
irrelevant, as long as at least one lookup table of corrections is
provided. In the end, for most of our frames, the automatic selection
implied the use of either a ``penny'' or a "moffat" function, and in
this case our tests indicate that the choice of one or another yields
offsets smaller than $5 \times 10^{-3}$~pix, again smaller than the
``repeatability'' of our measurements (see below), and therefore it is
of no concern.

\begin{figure}
\epsscale{.80}
\plotone{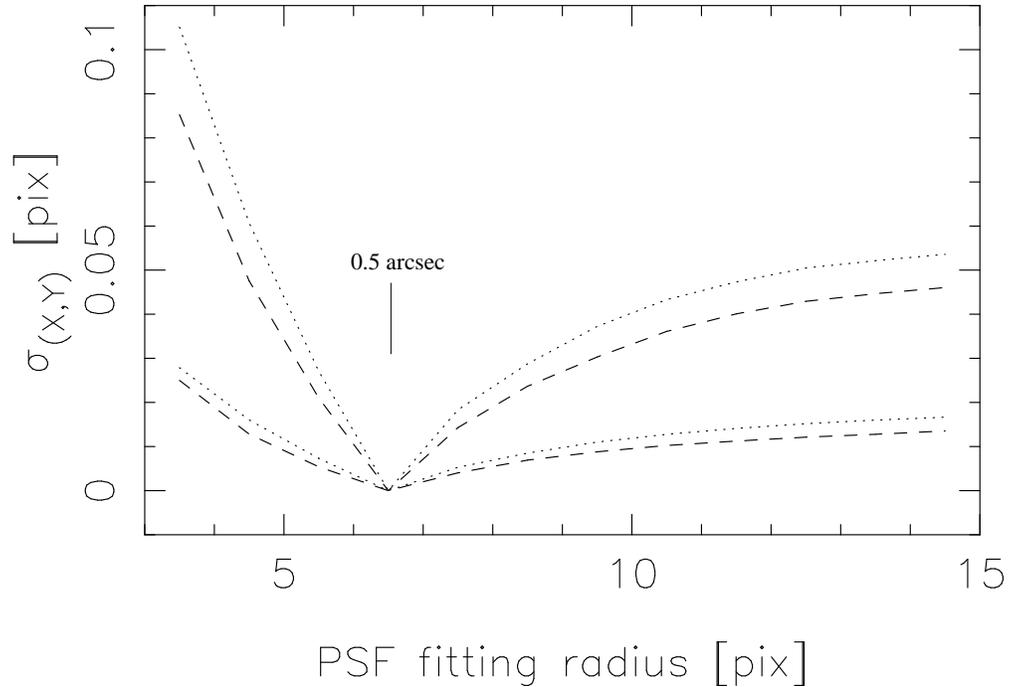}
\caption{Rms of coordinates (dashed lines = X-coordinate, dotted lines
  = Y-coordinate), derived from the same frame, as a function of the
  PSF fitting radius. The basic comparison (with $rms=0$, hence the
  sharp minima seen in the figure) is set at a FWHM of 6.5~pix,
  equivalent to the seeing of the analyzed frame, CTB-1 in
  Table~\ref{frames} (0.5~arcsec). The upper curves are for the entire
  sample on Figure~\ref{psftrends2} (i.e., $R \le 23.2$), whereas the
  lower curve is for the ``well-measured'' sample with $R < 21$,
  equivalent to our LRS stars). \label{fitrad1}}
\end{figure}

Besides the actual PSF function adopted, there are various other
parameters that control the way in which the PSF is used to determine
centroids on the image, e.g., the width and radius of the sky annulus,
the radius over which the of psf model is constructed, the actual PSF
fitting radius used to determine centroids. Our tests indicate that,
of all these parameters, it is the latter that has the most important
relative influence on the derived centroids. In Figure~\ref{fitrad1}
we show the rms scatter of coordinates from the same frame, CTB-1, as
determined by the same (automatic) PSF model as a funtion of the
adopted fitting radius. For this frame, with a seeing of 0.5~arcsec,
the usually recommended value for crowded field photometry
\citep{mas92}\footnote{Document available at
  http:$/$$/$iraf.noao.edu$/$docs$/$photom.html} equal to the FWHM (=6.5~pix)
has of course an rms with respect to itself of zero. The upper curves
are for the entire sample of stars, while the lower curves are for the
brighter portion of the sample (more representative of the LRS
stars). The smooth behavior of the (lower) curves in
Figure~\ref{fitrad1} allows us to infer that the {\it exact} value of
the adopted PSF fitting radius does not have an important impact on
the stability of the derived coordinates but, of course, being a
comparison of the coordinates derived from the same frame, it does not
tell us about the {\it optimum} PSF fitting radius to use.

\begin{figure}
\epsscale{1.1}
\plotone{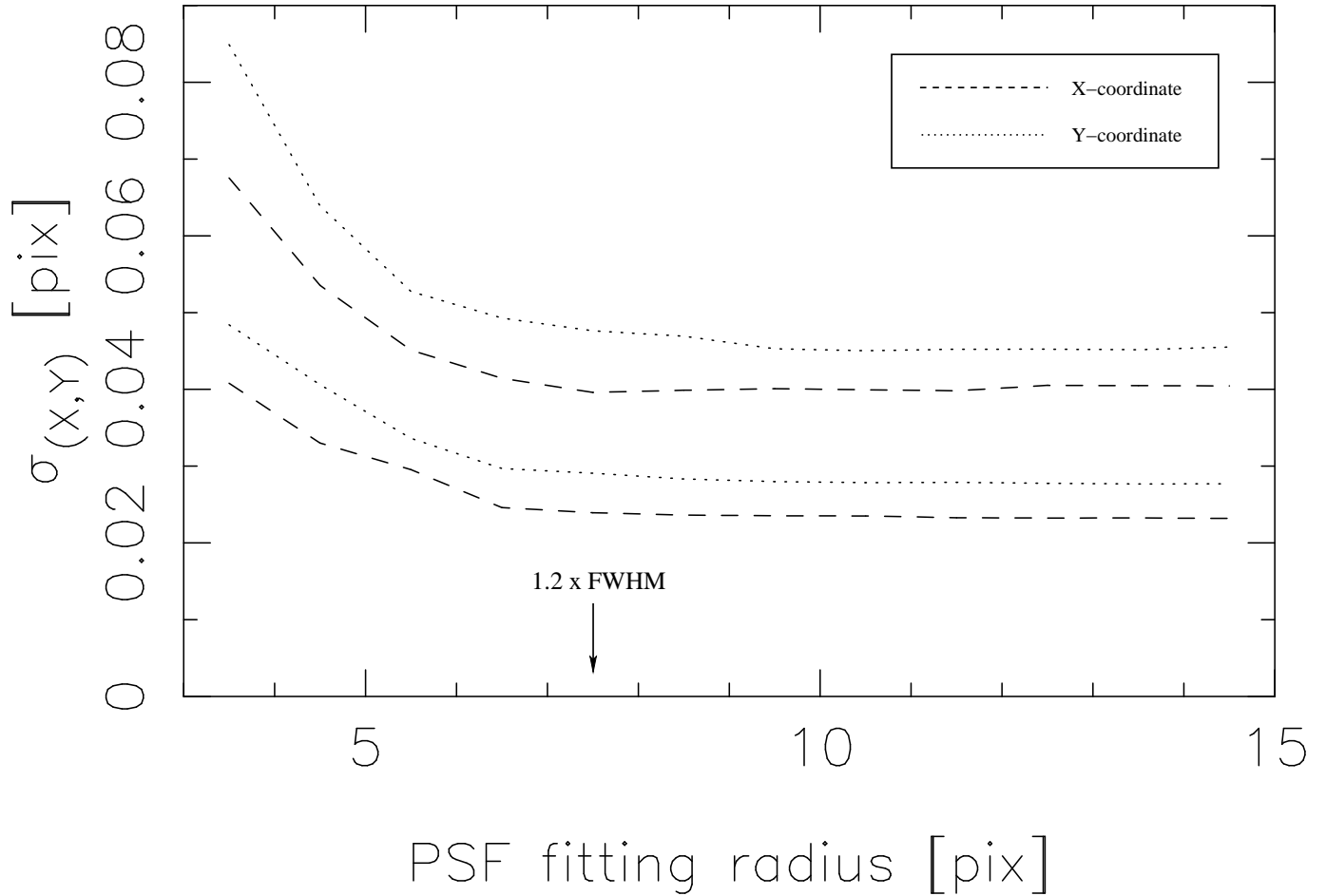}
\caption{Rms of coordinates (dashed lines = X-coordinate, dotted lines
  = Y-coordinate), derived from two consecutive low-$|HA|$,
  good-seeing frames (CTB-1 and CTB-2 in Table~\ref{frames}), as a
  function of the PSF fitting radius. The upper curves are for a
  sample similar to our LRS system (i.e., $R \le 23.2$), using a
  simple linear registration between the two frames, while the lower
  curve uses a full-quadratic registration and a slightly brighter
  sample. Both sets of curves show that a value slightly larger than
  the FWHM should be preferred for the PSF fitting radius. The value
  adopted throughout this work ($1.2\times$FWHM = 7.5~pix = 0.6
  arc-sec in this case) is indicated.\label{fitrad2}}
\end{figure}

To actually estimate our centering accuracy, and the best PSF fitting
parameters to use, we compared the coordinates derived from a pair of
the best FWHM consecutive frames, CTB-1 and CTB-2 in
Table~\ref{frames}. After determining an independent PSF for each
frame (but using the very same PSF stars and PSF parameters for both
frames), we computed $(x,y)$ coordinates for all our stars, and
computed the rms of the pair of coordinates as derived from these two
frames, as a function of the PSF fitting radius. Of course, the
expectation is that best fitting radius should yield the smallest
scatter on our coordinates. A small complication arose (see next
paragraph) in that we had to register the coordinates, which
introduced an additional source of noise (which, however, does not
affect these conclusions): Once we had determined an optimal
registration, the parameters from this unique transformation was used
for the coordinates derived from all pairs. As it can be seen from
Figure~\ref{fitrad2}, a value for the fitting radius slightly larger
than the FWHM of the image should be preferred. For both sets of
curves in Figure~\ref{fitrad2}, the rms decreases by about 5\% when
increasing the PSF fitting radius from the (nominal) FWHM to
1.5$\times$FWHM. Of course, when augmenting the fitting radius one
runs into the increased risk that the centroids become more
susceptible to possible faint nearby objects. For this reason, and as
a compromise value, throughout this work we adopted a PSF fitting
radius always equal to 1.2$\times$FWHM. Another important point from
Figure~\ref{fitrad2} is the decrease of almost a factor of two in the
registration residuals when going from a linear to a full quadratic
registration polinomial (this is further explored in
Section~\ref{anal}, see also Figure~\ref{coef}). Finally, the behavior
and significance of the linear terms, which would address the
commonality of the $(X,Y)$ pixel scales, is also presented in
Section~\ref{anal}.

\begin{figure}
\epsscale{.80}
\plotone{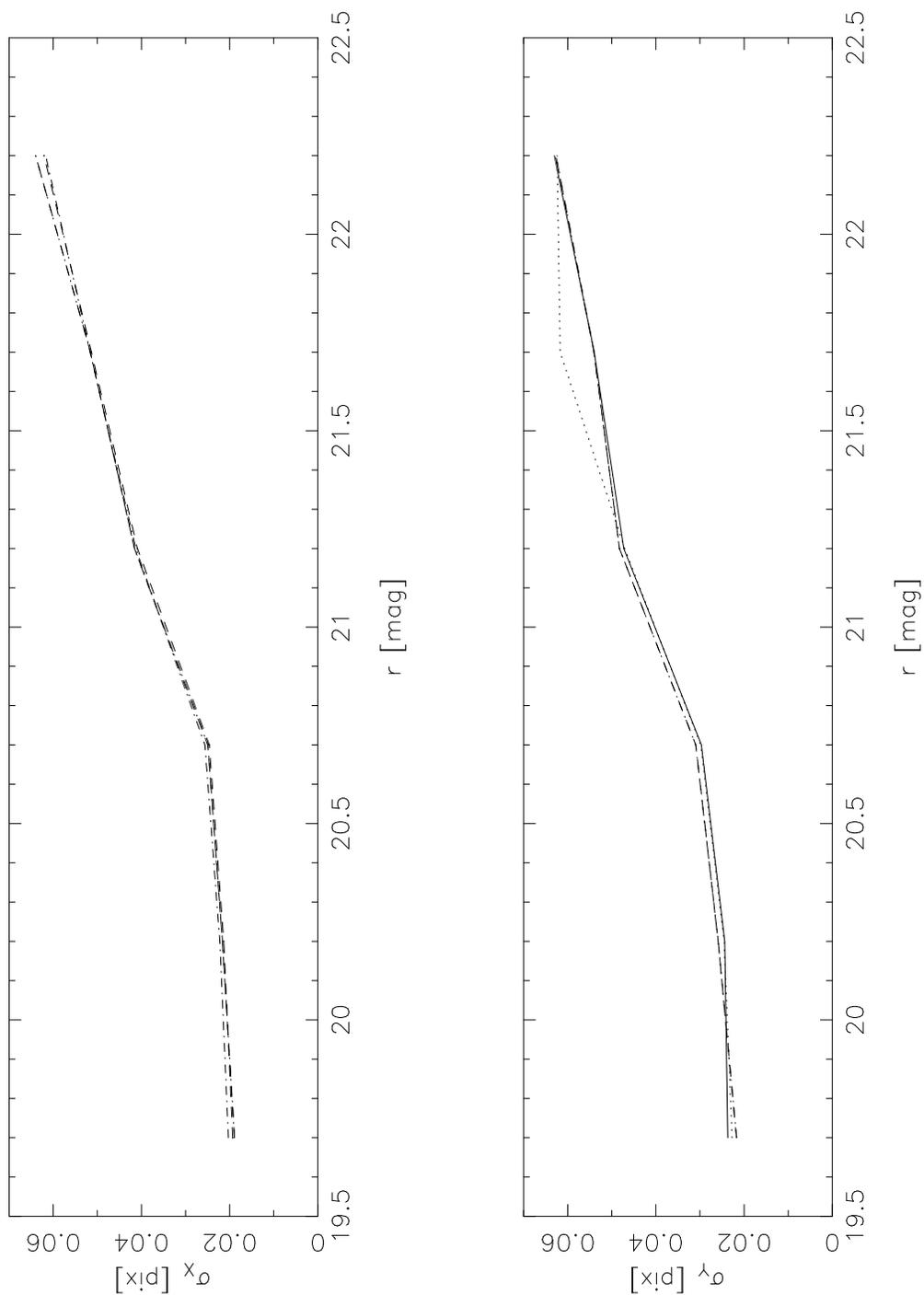}
\caption{Rms of coordinates in the X-coordinate (upper panel) and
  Y-coordinate (lower panel), derived from two consecutive
  small-$|HA|$, good-seeing frames (CTB-1 and CTB-2 in
  Table~\ref{frames}), as a function of the PSF R-band magnitude for
  (quadratic) transformation equations obtained using different
  magnitude cuts (and with correspondingly different rms residuals for
  the transformation). Dashed line is using all stars down to the same
  magnitude as the PSF magnitude to compute the transformation
  equations, solid line is for stars with $R< 20.7$ (transformation
  residuals of 0.027~pix per coordinate, 180 stars used) similar to
  the LRS stars), dotted is for $R< 22.2$ (transformation residuals of
  0.061~pix per coordinate, 730 stars used), and dot-dashed for $R<
  19.7$ (transformation residuals of 0.020~pix per coordinate, 67
  stars used). \label{transf}}
\end{figure}

\begin{figure}
\epsscale{.80}
\plotone{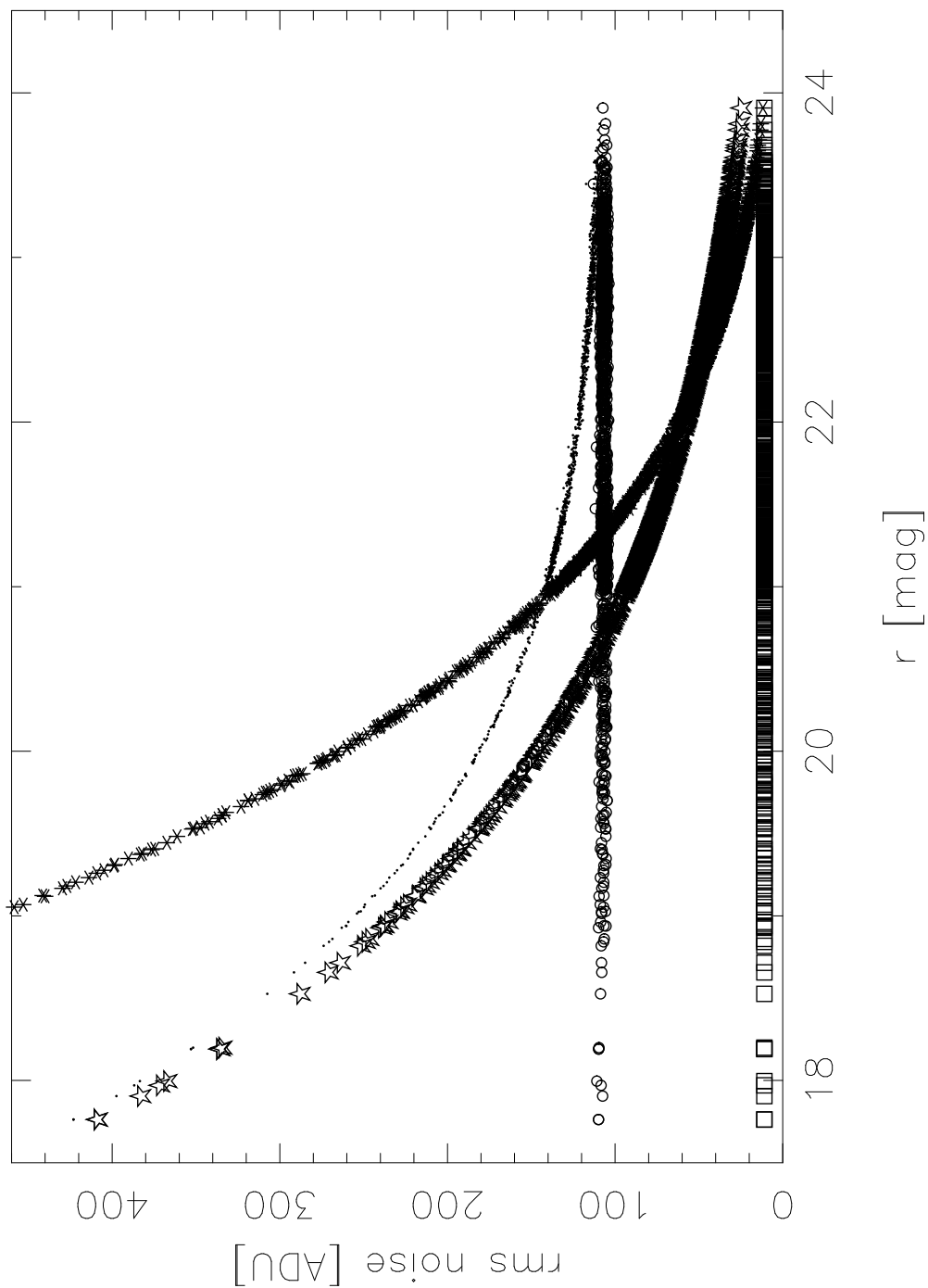}
\caption{Various rms noise sources {\it vs.} PSF r-band magnitude as
  measured on CTB-1 within an aperture of 6.5 pixels. Open star, open
  circle, and open square symbols represent source, sky, and detector
  (Poisson) noise respectively. Detector noise is negligible in all
  cases. The small dots indicate the total noise and the asteriscs
  represent the overall $S/N$ (on an aperture radius of $2 times
  FWHM$) for each detected object. The ``cross-over'' when the source
  noise becomes comparable to the sky noise which indicates when our
  centering accuracy starts deteriorating rapidly, happens for a $S/N
  \sim 200$. \label{sn}}
\end{figure}

As it is clear from the discussion in the previous paragraph, and also
from Figure~\ref{fitrad2}, our estimates of the unit weight
measurement error in the centroids is somewhat dependent on how well
determined is the registration between the frames we are comparing. To
study this particular point, we computed the rms of the coordinates by
using various schemes of transformation equations. The most relevant
conclusions from these tests is that the geometric transformation
equations (registration) derived from data sets with different (faint)
mag cuts (and with correspondingly different overall rms in the
transformation itself), do not have an important impact in the derived
$\sigma_{\mbox{X,Y}}$ {\it vs.}  magnitude (see
Figure~\ref{transf}). This means that the rms of the coordinates
represents their intrinsic centering scatter, and {\it not} the effect
of uncertainties in the registration itself. From this we conclude
that, e.g., for ``well measured'' stars with $R < 20.7$ (i.e., similar
to the mag cut adopted for our LRS stars) the rms of the pixel
differences is about 0.025~pix, or 0.018~pix ($\sim1.5$~mas) pix for a
single measurement, per coordinate. We also note that the sufficiently
large number of stars with this magnitude cut ensures a good X-Y
distribution over the entire FOV of the detector to map out the
geometric distortions. We finally note the apparent "break" of rapidly
increasing scatter for $R > 20.8$. As it can be appreciated from
Figure~\ref{sn}, this is when, roughly, the source noise becomes
comparable to the sky noise: In general, both throughout this work as
well as in previous works from our group (e.g., \citet{cos09}) we
adopt the magnitude for which $S/N > 200$ as a guide for the faint cut
of the LRS stars.



After having obtained the raw pixel coordinates, the subsequent steps,
all the way to obtaining PMs, follows very closely the procedure
described by \citet{cos09},
and includes computing LRS barycentric coordinates for each frame,
computing the differential chromatic constants in this barycentric
system (see next paragraph) for each LRS star and the QSO (using a
series of frames taken specifically for this purpose (see
Table~\ref{frames})), and using these constants to correct the
coordinates, registering all the LRS barycentric DCR-corrected
coordinates onto a unique astrometric reference system (the SFR), and
finally computing PMs for all LRS stars and the QSO through a simple
linear fit of the barycentric DCR-corrected \& registered coordinates
{\it vs.}  epoch such that, in the end, what we really measure is the
reflex motion of the QSO relative to the bulk motion of star field
(see Figure~\ref{pmall}), by changing the signs we obtain the motion
of the bona-fide Fornax stars with respect to the QSO, which thus
defines the zero point of the astrometric system). In the following
subsections we described in more detail than that given by
\citet{cos09} some aspects of these reduction steps.

We note that the field analyzed here actually contains two images of
the same QSO~J0240-3434, which is a known gravitational lens system
\citep{tin95}. Indeed one of the motivations to carry out this
analysis in this particular was field was that we could have two
independent measurements (using both images of the QSO) to assess our
final positional and PM precision. However, this attempt was
frustrated because one of the images (the NW image, component A),
happened to have a nearby (foreground) companion star (presumably a
member of Fornax), as shown in Figure~\ref{qsoab}. The companion was
nearly resolved in the best images available, but was hidden in the
lesser quality frames. Preliminary PM solutions using component A
turned out to be unstable (e.g., strongly dependent on the image
quality of the frames adopted in the solution), and with a large
scatter in the position {\it vs.} epoch diagram (the equivalent of
Figure~\ref{qsopm}). For this reason, these preliminary reductions
were discarded altogether, and image A was not used at all in our
astrometric solution. Therefore, all our results on this paper are
based exclusively on component B of QSO~J0240-3434.

\subsection{Pixel coordinates pre-corrections} \label{precor}

At this stage we are in a position to pre-correct our coordinates from
any known effect altering their true focal-plane position. Of course,
the first step would be to correct the coordinates from any known
optical distortions in the system. Unfortunately, no dedicated study
is available in the literature for the optical distortions on SuSI2
(as is the case of SOFI, another Nasmyth instrument at the
NTT~\footnote{http:$/$$/$www.obs.u\-bordeaux1.fr$/$m2a$/$ducourant$/$publications$/$distorsion\_SF.pdf},
and even this only deals with linear terms), nor could we, in the
allocated time, to obtain images of a suitable astrometric calibration
field to carry out such a study. We therefore do not apply any
pre-correction to our raw coordinates at this stage, and deal with the
effects of optical distortions through a (polynomial) registration to
be applied later on. We note that, as long as the distortions are
stable in time, by placing all targets at nearly the same location on
the CCD, it is the derivatives of the distortions that matter, and
these are likely to be very small (although not negligible, see
Section~\ref{anal}).



The next relevant pre-correction to our coordinates comes from
atmospheric refraction. For this we followed the prescription given by
\citet{sto96} with some modifications described in what
follows. First, we implemented an iterative procedure to get true
values starting from observed values, since atmospheric refraction is
formulated in terms of true values (not knwon a priori). More
specifically, we know that the relationship between the observed
zenith distance ($z'$) and the true one ($z$, not affected by
refraction) is given by (see e.g., Equations~(5.5) and~(5.7) in
\citet{taf91}):

\begin{equation}
z - z' = R_m(z) \label{zen}
\end{equation}

where $R_m(z)$ is the refraction constant evaluated at the true zenith
distance $z$ (see \citet{sto96}). Since the correction to $z$ is small
for the zenith distances considered in this work ($z'< 50^o$), our
procedure starts by assuming that on the right-hand-side of
Equation~(\ref{zen}) we can approximate $z \sim z'$ which allows us to
compute (an approximate) $R_m(z)$, which then gives us $z$. By
successive iterations, we rapidly (three to four iterations) converge
into $z$ (and $R_m(z)$).

The computation of the index of refraction, related to $R_m(z)$ (see
Equations~4 and 5 on \citet{sto96}) requires knowledge of the
atmospheric temperature, pressure, and relative humidity at the
instant of data acquisition and the bandpass of the observations. All
these parameters are available on the image headers, and are read from
the observatory's weather station. Latitude ($-29^015'18.4''$), and
altitude (2,375~m above sea level) for the observatory were adopted
from the observatory's web
page\footnote{http:$/$$/$www.eso.org$/$sci$/$facilities$/$lasilla$/$telescopes$/$ntt$/$overview$/$techdetails.html}. Other
relevant parameters (observed RA \& DEC, LST) were also read from the
image headers. For the calculation of the index of refraction (see
Equations~14 to 17 in \citet{sto96}), we adopted the central
wavelength of the R-band filter, $6415.8\,\AA$, as given on the
filter-page for
SuSI2\footnote{http:$/$$/$www.eso.org$/$sci$/$facilities$/$lasilla$/$instruments$/$susi$/$docs$/$SUSIfilters.html}. With
all this information, the raw pixel coordinates were corrected for
refraction adopting the relevant ambient parameters for each
individual frame.

So far, in the previous step, we have evaluated the refraction
constant at one particular wavelength. However, the refraction
constant $R_m(z)$ actually depends on the spectral energy distribution
of the star being observed (see e.g., Equation~(22) in
\citet{sto96}). Because on our LRS we have stars of different spectral
types, as well as the QSO, the refraction constant will actually
differ for all of them. This gives rise to a ``differential
displacement'' of star positions, depending on their spectral energy
distribution, which is a function of $z$ (see Equation~(\ref{zen})),
known as Differential Chromatic Refraction (DCR). By contrast, the
refraction correction described in the previous section affects all
stars equally, regardless of their spectral type, and could be thus
called ``continuous refraction'' to differentiate it from DCR. To
correct our coordinates from DCR effects we followed the procedure
indicated in \citet{cos09} (Section~3.4), for which we give a more
detailed justification here. We must emphasize that these
corrections are {\it differential} in the strict sense that they are
computed with respect to the mean of the {\it current} LRS. More
specifically, the {\it computed} barycentric coordinate, e.g., in the
X-direction, for star ``$i$'', $X'_{i_b}$, is given by:

\begin{equation}
X'_{i_b} = X'_i - \frac{\sum_{j=1}^{N} X'_j}{N}  \label{bari1}
\end{equation}

where $X'_i$ are the pixel coordinates measured on the detector, and
$N$ is the (current) number of LRS stars that define the (current)
barycenter. The primed pixel coordinates\footnote{following the same
  notation as on \citet{cos09}}, being measured directly, are of
course affected by atmospheric refraction. The relationship between
the refraction-free coordinates $X_i$ (assumed to be oriented in the
direction of RA), and the ones affected by refraction can be computed
from Equation~\ref{zen}, and is given by Equation~(29) on
\citet{sto96}, which we include here for completeness:

\begin{equation}
 X'_i - X_i = R_{X_i} \times \sin \eta'   \label{refra}
\end{equation}

where $R_{X_i}$ is the refraction constant for star $i$,


and $\eta'$ is the so-called ``angle of the star'' (or parallactic
angle\footnote{See e.g., page 71 on Smarts (1962)}) given by
Equations~(3) and~(4) in \citet{cos09}, which is itself a function of
$z'$ and the observed hour angle $H'$. Equation~(\ref{refra}) clearly
shows that, due to atmospheric refraction, $X'_i$ is a function of
$z'$ and $H'$, i.e., one can write $X'_i=X'_i(z',H')$.

Replacing $X'_i$ from Equation~(\ref{refra}) into
Equation~(\ref{bari1}), after some elementary algebra, we can see
that, for the barycentric coordinates, it is valid that:

\begin{equation}
X'_{i_b}(z',H') = X_{i,b} + \left( R_{X_i} \times \sin \eta' -
\frac{\sum_{j=1}^{N} R_{X_j} \times \sin \eta'}{N}
\right) \label{bari2}
\end{equation}

where, $X_{i,b}$ is the barycentric coordinate free from
refraction ($\equiv X_i - \frac{\sum_{j=1}^{N} X_j}{N}$). In
principle, the value of $\eta'$ inside the summation on the right hand
side of Equation~(\ref{bari2}) should be computed for the exact
position of (each) star $j$ in the LRS. However, given our small FOV,
this quantity can be considered constant, and therefore,
Equation~(\ref{bari2}) can be re-written as:

\begin{equation}
X'_{i_b}(z',H') = X_{i,b} + \left( R_{X_i} - \frac{\sum_{j=1}^{N}
  R_{X_j}}{N} \right) \times \sin \eta' \equiv X_{i,b} + R^{DCR}_{X_i}
\times \sin \eta' \label{dcr}
\end{equation}

Where we have re-defined the (classical) refraction constant to a new
DCR constant given by $R^{DCR}_{X_i} \equiv R_{X_i} -
\frac{\sum_{j=1}^{N}}{N}$. Equation~(\ref{dcr}) reveals the true
nature of the DCR correction, $R^{DCR}_{X_i}$ is nothing but a
refraction constant for each LRS object with respect to the mean of
the refraction constants for the (current) LRS, hence its name
``differential''. A completely equivalent equation can be written for
the Y ($\sim$DEC)-coordinate.

As explained by \citet{sto96} (see, e.g., his Equation~(3)) one can
develop $R_m(z)$ as a power series of $\tan z$. For small zenith
distances, such as those involved here, one can retain the first order
dependence of $R_m$ on $\tan z$ (i.e., $R_m \sim R'_m \times \tan z'$, see,
e.g., Equation~(5.7) on \citet{taf91}), and in this case
Equation~\ref{dcr} can be written as:

\begin{equation}
X'_{i_b}(z',H') = X_{i,b} + \left( R'_{X_i} - \frac{\sum_{j=1}^{N}
  R'_{X_j}}{N} \right) \times \tan z' \sin \eta' \equiv X_{i,b} + R'^{DCR}_{X_i}
\times \tan z' \sin \eta' \label{dcr2}
\end{equation}

As it was done in \citet{cos09}, a series of ``refraction frames''
with different values of $(z',H')$ were acquired for this field (see
Table~\ref{frames}). From this series, we then fitted the computed
(measured) barycentric positions of all our LRS stars (the left hand
side of Equation~(\ref{dcr2})) as determined from each DCR frame, as a
function of $\tan z' \sin \eta'$, and we computed the straight line
fit implied by the right hand side of Equation~(\ref{dcr2}), which
allows us to solve for the differential refraction-free coordinates
$X_{i,b}$ as the intercept for abscissa zero of that line. An example
of this fit, for the QSO and two anonymous stars from the LRS is shown
in Figure~\ref{dcrline}. This figure can be compared to its
equivalent, presented by \citet{cos09} (their Figure~4), for our PM
study of the SMC.

\subsection{From pixel coordinates to proper motions; The ``QSO method''} \label{method}

Once the LRS coordinates have been corrected for DCR in the way
described in the previous section, we can determine the PMs of all our
LRS stars and the QSO. The first step at this stage is to build up the
SFR mentioned in subsection~\ref{building}. The SFR is constructed
from the mean DCR-corrected barycentric coordinates using all the
initial LRS stars from the three consecutive, good quality,
low-$|HA|$, intermediate-epoch frames indicated as ``SFR'' in
Table~\ref{frames}. We then use a bi-dimensional $(X,Y)$ polynomial
fitting routine to compute, based on a $\chi^2$ minimization
algorithm, transformation equations (coefficients) that map the
barycentric DCR-corrected coordinates for the LRS stars (excluding the
QSO) from all the frames onto this SFR, and apply these coefficients
to transform the coordinates for all the LRS stars and the QSO into
the SFR, in the way described by \citet{cos09} (specially their
Section~3.5). Once the coordinates from all frames haven been
transformed into a common reference system (provided by the initial
set of 260 LRS stars on the SFR), we perform a simple linear $\chi^2$
fit of these coordinates {\it vs.} epoch based on all
good available frames (see Figure~\ref{qsopm}). We end up at this
stage with PMs for all the LRS stars, and the QSO. In
Figure~\ref{pmall} we show the vector point diagram based on all 260
initial LRS stars (see below), as well as for the QSO.

In the method outlined, the underlying assumption is that LRS stars
share a common PM so that their coordinates can be registered with
high confidence (to the extent of our astrometric accuracy) into a
common reference system. The reflex motion of the QSO is measured with
respect to this homogenous (kinematically speaking) set of objects, in
what is called the ``QSO method''. In our case, the expectation is
that most of the LRS stars are Fornax members, so the reflex motion of
the QSO with respect to the bulk motion of the Fornax stars is nothing
but the motion of the galaxy itself with respect to the QSO, which is
actually ``at rest'' (in the tangential direction, which is what
concerns us). We cannot however exclude the possibility that a few
Galactic foreground stars may be mixed in the LRS. Fortunately, since
we have determined PMs for all of the LRS stars, any Galactic star
will exhibit a motion different from that of the bulk of Fornax
stars\footnote{unless, of course, by chance, their motions is similar
  to that of Fornax}. We can therefore start eliminating objects with
spuriously high PM, indicative of them not belonging to Fornax. As can
be seen form Figure~\ref{pmall} there are actually no objects with
extremely large PM: There are only two objects with $\mu >
3.0$~$\masy$, and no objects with $\mu > 4.0$~$\masy$. We note that a
typical Galactic halo star would have a PM of $3.0$~$\masy$ at a
distance of 11~kpc. A rather conservative PM cut at $2.0$~$\masy$
would avoid any halo star out to a Heliocentric distance of 16~kpc
equivalent, for the Galactic position of Fornax ($l=237^o.1,
b=-65^o.7$) to a distance of $\sim$15~kpc from the Galactic
plane. Given the rather large Galactic latitude of our field, the
small FOV, and the magnitude and color range of our selected stars, we
expect a rather small Galactic contamination. Indeed, using the
Galactic star counts model described by \citet{men96}, we predict that
there is less than 1 Galactic star per square degree in the range $18
\le R \le 21.8$ and $0.5 \le B-R \le 2.5$ (see Figure~\ref{cmd1})
towards the Fornax field. For our precise FOV, we predict in total 11
Galactic stars satisfying our magnitude \& color cuts.

Following the above discussion, we identified 14 LRS stars that
satisfy $\mu>2.0$~$\masy$, and were thus eliminated from the initial
LRS, and the full registration (and PMs) for the remaining LRS stars
and the QSO re-computed. For comparison purposes, the intermediate PM
results for \qso\ from these different samples of LRS stars (and for
the conditions discussed in the next section) are given in
Table~\ref{pms}. As it can be seen, the elimination of these
relatively large PM objects did not modify our results
substantially. Furthermore, another 7 stars were eliminated from the
LRS because they exhibited residuals beyond $3\sigma$ of the formal
rms of the registration in at least 4 or more astrometric frames,
again the impact of their removal (see Table~\ref{pms}) was minimal.

The PM solution described so far used a full 3$^{rd}$ degree
polynomial in $(X,Y)$ for the registration, and include only the 15
frames satisfying that their $|HA|< 1.5$~h, as listed in
Table~\ref{frames}, in close resemblance to the strategy adopted in
\citet{cos09}. In the next section we analyze in more detail the
stability of our PM solution in regards to these (and other) specific
choices.

\newpage
\section{Analysis} \label{anal}

In Figures~\ref{cmd1} and~\ref{errpm1} we show the CMD for the 239
stars of the LRS selected as described in the previous paragraph, as
well as the QSO. Throughout this entire paper, our PSF photometry has
been approximately calibrated by using a photometric zero point
defined by the blue magnitude of the QSO ($B_{\mbox{QSO}}=19.94$) as
reported by \citet{tin97}. Unfortunately these authors did not report
any $R$-band photometry, which was instead scaled by adopting a color
for the red-clump on Fornax of $B-R \sim 1.3$, as implied by the deep
Fornax photometry from \citet{ste97}. Our CMD (plotted for an easy
comparison in the same color window as Figure~7 by \citet{ste97})
indicates that Galactic contamination seems to be minimal. In any
case, and in order to be absolutely safe and to test the sensitivity
of our method to these (possible) outliers, we eliminated the two
brightest \& reddest objects ($B-R \sim 2.3$), the one object below
the Fornax RGB/AGB sequence at $R \sim 20.7$ and $B-R \sim 1.8$, the
two faintest \& bluest objects at $B-R \sim 0.75$, and finally another
two suspicious objects on the upper/left envelope of the AGB/RGB. On
the other hand, the PM error {\it vs.} magnitude plots
(Figure~\ref{errpm1}) show that there are a few outlier points that
can also be eliminated from the LRS. In this case, if we cut at
$\sigma_\mu > 0.5$~$\masy$ (4 stars), and also eliminate (somewhat
subjectively) 11 objects that have unusually large PM error (in either
coordinate) for their magnitude (left upper \& lower panel on
Figure~\ref{errpm1}), we end up with an LRS sample of 217 stars. The
PM solution for this sample, indicated in the 4$^{th}$ row on
Table~\ref{pms} shows a very small variation with respect to our
previous solutions. The variation of the (DCR-corrected) barycentric
coordinates for \qso\ based on this sample of LRS stars in shown in
Figure~\ref{qsopm}. From now on, this solution will become our
``reference value'' to carry out the comparisons that follow. Another
interesting point brought to mind by Figure~\ref{errpm1} is that the
QSO shows a perfect agreement with the general trend of PM error {\it
  vs.} magnitude, indicating that its positional uncertainty is no
different at all from that of stellar images of the same brightness.

Following the procedure outlined in \cite{cos09}, all previous PM
solutions were computed from frames restricted to being close to the
meridian, to avoid any possible unaccounted DCR-related
effects. Since, to the best of our ability, we have actually corrected
our coordinates of this effect, one may consider including frames
outside this range. Given the $|HA|$ distribution of frames in
Table~\ref{frames}, increasing the $HA$ window to 2 hours only
incorporates two more frames to our solution. While by doing this our
errors slightly {\it decrease}, the actual PM does not change
much. Only by increasing the $HA$ window to 3 hours from the meridian
we start to see a larger change in the PM, and a corresponding {\it
  increase} in the overall error of the fit (although the error in PM
remains almost constant). Indeed, while the overall rms of the fit in
RA and DEC are (0.88,1.27)~mas for the $|HA|<1.5$~hrs solution (see
Figure~\ref{qsopm}), this increases to (0.92,1.36)~mas for the
$|HA|<3.0$~hrs solution. For these reasons, we adopt our original
solution, based on the 15 frames with $|HA|<1.5$~hrs. DCR effects are
often mentioned in the literature as playing an important role in
high-precision ground-based relative astrometry. To illustrate this in
the case of our data, we have computed a solution with $|HA|<1.5$~hrs
but {\it without any} DCR correction (i.e, straight raw pixel
coordinates). As we can see on Table~\ref{pms} the PM values do change
somewhat (but, completely within the uncertainties), and the PM errors
are slightly increased. In a way this is an expected result, since the
data from which the PMs are computed is quite close to the meridian,
and therefore the net DCR effect on the motions is not that relevant.

Our coordinates were also pre-corrected for continuous refraction as
described in Section~\ref{precor}. It can be seen from
Equation~(\ref{zen}) that since this correction affects only the
altitude (not the azimuth) of stellar positions, a failure to correct
for this effect could introduce a differential change of scale in the
$X$ and $Y$ directions (see Equations~(1) and~(2) in
\citet{cos09}). The correction is however small, and it increases with
HA of the frame. For example, for the most extreme DCR frame at
HA=-4~h (see Table~\ref{frames}), the registration into the SFR frame
yields a coefficient for the X-term (which would give the differential
scale in the x-coordinate) that differs from 1.0 (= equal scale
between the two frames) by $\sim 10^{-4}$ if we pre-correct for
continuous refraction. If we do not pre-correct for refraction, the
difference to 1.0 becomes $6 \times 10^{-4}$. In the y-coordinate the
equivalent numbers are $\sim 10^{-4}$ and $7 \times 10^{-4}$
respectively. Therefore, the continuous refraction pre-correction {\it
  does} help in rendering the scales more equal (for a given
coordinate) for all frames, irrespective of their zenith
distance. Another point is that the {\it difference} in the x- and y-
plate scales is $\sim 10^{-4}$ for the refraction correction
coordinates, while this number increases to $5 \times 10^{-4}$ if no
refraction is applied. Therefore continuous refraction also renders
plate scales more equal in both coordinates. We finally note that our
registration procedure seems to be able to account for this effect
even if no pre-correction is applied, for example, the error in the x-
and y- coefficients for the example just described is the same for the
corrected and non-corrected data ($\sim 1.5 \times 10^{-5}$). In
general, if the required ambient data is available, pre-correction for
continuous refraction is recommended (William van Altena, private
communication).

A critical step in the whole process of obtaining the PMs is to
adequately mapping the coordinates from different epochs into the
SFR. The SFR itself is built from as many as possible (in our case 3,
see Table~\ref{frames}) good-seeing contiguous frames: The average of
their barycentric coordinates provides, in principle, a more stable
reference system than that of any individual frame, subject to the
centroiding errors discussed in section~\ref{centroids}. Therefore,
the SFR better represents the true geometry of the FOV, and allows us
to properly map out any offset, rotation, change of scale and, in
general, higher order optical distortions between frames of different
epochs and the SFR itself. Since we are only concerned with {\it
  relative} displacements as a function of time, by placing the QSO
(and of course by extension all the LRS stars) always close to a
fiducial reference point near the center of the chip as discussed in
section~\ref{obma}, we minimize the relative distortion from different
sections of the chip. In our standard procedure \citep{cos09}, we take
a straight average of the SFR frames, without any registration between
them: This might seem justified by the fact that they are contiguous
good-quality frames. However, if one computes the scatter of the
coordinates for the three frames that contribute to our SFR, we have
an rms of $(\sigma_x,\sigma_y=(6.1,2.7) \times 10^{-2}$~pix for the
straight average, while this values decreases almost a factor of ten
if we do register (quadratically) the SFR frames:
$(\sigma_x,\sigma_y=(6.8,6.0) \times 10^{-3}$~pix (in both cases the
computed rms is for all the stars brighter than $R~\sim 20.7$
discussed in section~\ref{centroids}). This prompted us to test
whether our PM results were sensitive to our straight average of the
frames that conform the SFR. For this purpose we created another SFR
this time with the three frames that make it registered into one of
them by means of a cubic polynomial, and computed our PMs again. The
results for the QSO PM (and its error) were equivalent. Therefore, to
the level of accuracy of our registration (see next paragraphs), the
use of either SFR is completely appropriate.

The classical way to discern on the proper modeling of the
registration (besides comparing to an external catalog with
positions of much higher accuracy than our own - not available in this
case) is to look for position-dependent residuals, after registration,
as a function of the various relevant positional \& flux parameters,
and to choose the lowest order correction that, within the
uncertainties, leaves no correlations (see, e.g., Figures~9, 10 and~11
in \citet{cos09}). In our standard procedure (following \citet{cos09})
we have used a full 3$^{rd}$ order polynomial fit. On the other hand,
the PM results, using our ``clean'' sample of 217 LRS stars when using
either a $2^{nd}$ and a $4^{th}$ order polynomial exhibit rather large
variations, see Table~\ref{pms}. We can however conclude that, since
the PMs are within $1\sigma$ from each other when comparing our
3$^{rd}$ {\it vs.}  our $4^{th}$ order polynomial, we would prefer the
lower order registration of these two. Deciding between the $2^{nd}$
and 3$^{rd}$ order registration is less straightforward. We note that
in {\it both} cases the registration yields rms positional residuals
with respect to the LRS of 0.018~pix (as already found in
section~\ref{centroids}), so one might tend, again, to choose the
lower $2^{nd}$ order polynomial. We have however reasons to prefer the
3$^{rd}$ order registration: First, the rms of the barycentric
displacement of the QSO has an overall error 5\% larger in the
$2^{nd}$ order registration ((0.97,1.21)~mas in (RA,DEC) respectively)
with respect to the 3$^{rd}$ order registration
((0.88,1.27)~mas). Secondly, a plot of the registration coefficients
indicates that, at least some of the 3$^{rd}$ degree coefficients are
very significant: In Figure~\ref{coef} we plot, in logarithmic scale,
what we could call the ``significance'' of each coefficient (given by
$\log_{10}(\mbox{Coefficient}/\sigma_{\mbox{\small{coefficient}}})$ of a
3$^{rd}$ order registration polynomial. As we can see, in the
X-coordinate, the only non-significant term seems to be the $Y^3$
coefficient, while in the Y-direction, one could perhaps discard the
$X^3$ and the $XY^2$. All other terms appear significant. Even though
the specific terms just mentioned seem irrelevant, their inclusion
does not hurt our solution: The QSO PM solution leaving these 3$^{rd}$
order terms out of the registration polynomial produces a solution
that is indistinguishable from the full 3$^{rd}$ order registration
(see Table~\ref{pms}), which we then adopt as our standard. An example
plot of the registration residuals is shown in Figure~\ref{resi}, as
mentioned before the rms of this registration is 0.018~pix.



One other aspect that can seriously affect the registration is the
number and distribution of LRS stars. Our LRS stars are
randomly distributed all over the field (see Figure~\ref{xylrs}), but
one may wonder whether their number is adequate. If one takes the
usual rule-of-thumb that for a $\chi^2$ polynomial fit of $N$
coefficients one needs at least $3 \times N$ data points, then our
number of LRS stars (217) should be plenty to fit our required 10
coefficients per coordinate (see Figure~\ref{coef}). To test the
sensitivity of our results to the actual number of LRS stars we
performed a solution deleting one every-other entry in the LRS list of
stars, ending with 109 LRS stars. The PM solution in this case (see
Table~\ref{pms}) indicates a very small change of our results (less
than $1\sigma$), showing what we could call a ``stable'' LRS (i.e. our
results are independent of which Fornax stars conform the LRS).

In our method, we calculate PMs for both the QSO (with respect to the
motion of the LRS stars), as well as for each LRS star (with respect
to the mean of their motion). If all LRS objects are genuine Fornax
stars, all these latter motions are not ``real'' in the sense of not
representing any true kinematic signature of Fornax stars, rather they
are indicative of our internal errors. Indeed, for a velocity
dispersion of a few $\kms$ (typical of the dSphs), this is equivalent,
at the distance of 138~kpc \citep{mat98}, to a proper motion of $\sim
8 \mu$as~y$^{-1}$, much below our measurement
uncertainties. Therefore, the ``dispersion'' of points in the cloud of
LRS stars in Figure~\ref{pmall} does not represent true proper
motions, but what we could perhaps call ``displacements'' (or
pseudo-PMs, or ``residual motions'' as in \citet{cos09}) due to our
various positional measurement uncertainties. For this reason we
believe that the use of these displacements to enter into an iterative
process to pre-correct the coordinates at different epochs by using
the derived pseudo-PMs, and in this way improve the quality of the
registration, is not a recommended strategy in this particular case,
and we have not implemented it.

In Figures~\ref{pmxy} and~\ref{pmmag} we plot the (pseudo-)PM for our
final sample of 217 LRS stars, and the QSO as a function of position
and photometry. As it can be seen from these figures, no significant
trends are found. We thus believe that our motions are not affected,
as far as we can measure them, by any obvious systematic effects that
depend on these parameters.

\section{Comparison to other studies \& Conclusions} \label{comp}

There have been only two astrometric determinations of the PM for the
Fornax dSph galaxy, namely that by \citet{din04}, based on a
combination of ground-based plates and Hubble-WFPC data, and that
based exclusively on HST data (\citet{pia07}, which gives revised
values to those reported earlier in \citet{pia02}).

\citet{din04}, based on an independent determination using 48 galaxies
and 8 QSOs give a weighted mean of (\mua,\mud)=($0.59 \pm 0.16, -0.15
\pm 0.16$)~\masy. On the other hand, \citet{pia07}, based on 4 QSOs
(their Table~3), gives a weighted mean of (\mua,\mud)=($0.476 \pm
0.046, -0.360 \pm 0.041$)~\masy. In order to compare our result (based
on a single QSO) with their measurements it makes more sense to compare
our value with their {\it individual} PM values. This allows us to
estimate the expected uncertainty on our final weighted mean when we
incorporate our other 4 QSO fields. In Figure~\ref{pmcomp} we plot the
individual measurements from these previous works, along with our
measurement. We note however that, while \citet{pia07} report
individual measurements, \citet{din04} only give the mean values with
respect to 8 QSOs and 48 galaxies, and these are the results plotted
in Figure~\ref{pmcomp}.
\citet{din04} {\it do} give the {\it individual} PMs for images A and
B of the same QSO reported in this paper. These individual values
are also plotted in Figure~\ref{pmcomp}. By looking at the open and
filled squares we can see the improvement achieved by ground-based
astrometry when using CCDs and a homogeneous data set in comparison
with the results that combine non-linear plates and heterogeneous
data. Nevertheless, it is somewhat surprising how close our (single)
value is to the mean PM from galaxies as derived by \cite{din04}
(differences smaller than $1\sigma$ of {\it our} smaller error), while
there is a larger discrepancy (although still smaller than $2\sigma$
of {\it their} (larger) error) when compared to the PM using QSOs. In
Dinescu's study they have a large sample of brighter better-measured
galaxies (see their Figure~2) while the (much fewer) QSOs are fainter
and possibly have larger positional errors (although, their individual
PM errors are smaller than those of galaxies for a given magnitude all
the way down to V$\sim 21$, possibly due to the extended, more diffuse
nature of galaxies). Overall, their mean PM with respect to galaxies
has an error $\sim50$\% smaller than that derived from the QSOs.

\begin{figure}
\epsscale{.80}
\plotone{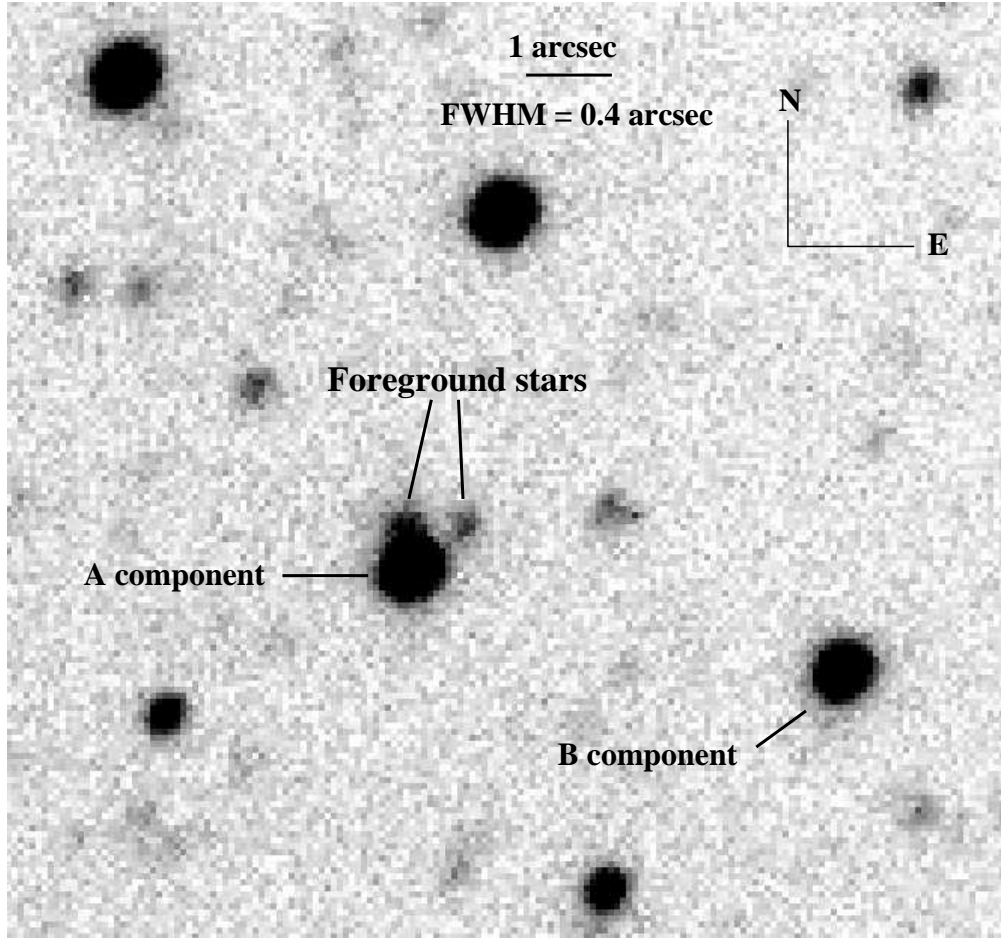}
\caption{Best quality image (frame acquired on 2000-08-08, see
  Table~\ref{frames}) with a FWHM=0.4~arcsec, indicating the position
  of the A and B components of field QSO J0240-3434. The A component
  has two nearby foreground stellar companions (possibly members of
  Fornax). These stellar images blend with component A in the lesser
  quality frames of other epochs, deteriorating the astrometry. Only
  the SE (B) component was used throughout this work. Image size is
  $\sim$12~arc-sec on a side, and the image scaling is linear in
  ADUs. Several other foreground Fornax stars appear in the
  FOV. \label{qsoab} }
\end{figure}

\begin{figure}
\epsscale{1.2}
\plotone{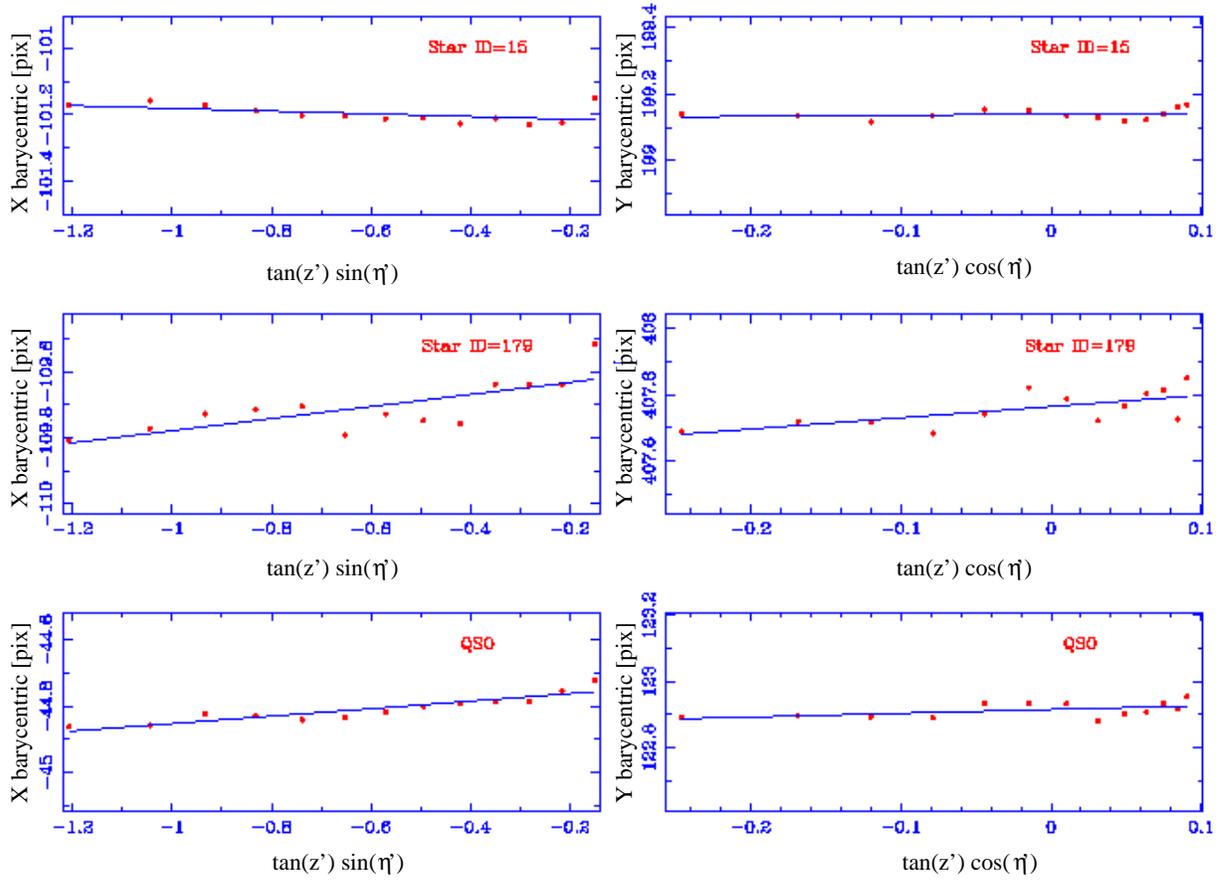}
\caption{Example DCR series plots for two randomly selected reference
  stars, and the background QSO J0240-3434B. They are based on a series of 13
  off-meridian consecutive frames labeled as ``DCR'' in
  Table~\ref{frames}. \label{dcrline} }
\end{figure}

In comparison with the individual HST measurements, we see a rather
large discrepancy, specially in DEC: While our value lies between 2.3
and 1.2$\sigma$ away from HST individual PMs in \mua, this difference
becomes between 4.4 and 2.4$\sigma$ in \mud. We note that the largest
\& smallest PM values from HST for {\it both} \mua\ and \mud\ (see
Table~3 on \citet{pia07}) come precisely from images A and B of
\qso. \citet{din04} also point the rather large discrepancy in the PMs
derived from components A and B (plotted in our
Figure~\ref{pmcomp}). As was described in section~\ref{centroids} the
A component of QJ~0240-3434 {\it is} indeed affected by two close
companions (see Figure~\ref{qsoab}). It is not unlikely that tiny
changes in the photocenter of the images due to other close,
unresolved and fainter, companions, or even slightly extended
structure(s) in the wings of the QSO (from the underlying QSO galactic
disk) could introduce an extra source of noise in the PMs that is not
necessarily accounted for in the final PM error budget. Obviously,
whether or not our own measurements are subject to some (as yet
unknown) systematic effect related to what we just mentioned, is an
issue that could be addressed once we incorporate the other QSO
fields. In Table~\ref{pmtab} we summarize the measurements plotted in
Figure~\ref{pmcomp} (the ``perspective'' motions are explained further
below).

It is interesting to ascertain to what type of final (Heliocentric)
velocity errors correspond our individual PM measurement errors. For a
given distance and distance error $r \pm \sigma_r$ and PM and its
error in component ``x'' (in this case either RA or DEC), $\mu_x \pm
\sigma_{\mu_x}$, the corresponding velocity and its error is given by:

\begin{eqnarray}
v_x = K \, r \, \mu_x \label{vel} \\ \nonumber \\
\frac{\sigma_{v_x}}{v_x} = \sqrt{ \left( \frac{\sigma_r}{r} \right)^2 +
\left( \frac{\sigma_{\mu_x}}{\mu_x} \right)^2 } \label{velerr}
\end{eqnarray}

where, if $r$ is in kpc, and $\mu_x$ is in \masy, then $K=4.74$ and
$v_x$ and $\sigma_{v_x}$ are in km~s$^{-1}$. According to
\citet{mat98}, the distance to Fornax is $138 \pm 8$~kpc (i.e., 6\%
formal error). Therefore in equation~(\ref{velerr}) the distance error
is negligible in comparison with our PM errors in either RA or
DEC. For our PM reference value (see Table~\ref{pmtab}) we obtain a
Heliocentric velocity and error of $v_{\alpha}=419 \pm 58$~km~s$^{-1}$
and $v_{\delta}= -7 \pm 72$~km~s$^{-1}$. The measured Heliocentric
radial velocity from \citep{mat98} is $v_r = 53 \pm 3$~km~s$^{-1}$. It
is thus clear that the derived Heliocentric Fornax motion is dominated
by uncertainties in the PM from our one-QSO measurement, not by the
distance or radial velocity uncertainty. Given our measurement errors,
once we incorporate our other four QSO fields, we expect to achieve
velocity errors for the weighted mean on the order of 30~km~s$^{-1}$
per component, totally compatible with the (weighted mean) HST
results. Table~\ref{pmtab} also shows that our proper motion value
exhibits the largest tangential velocity for Fornax after the Dinescu
et. al value from Galaxies. At this stage we howevere refrain from
making a full analysis of the implications of this result, until we
have collected the data from the other four QSO fields which will
allow us to have a more robust weighted mean.

We must note that, when comparing PMs values derived for an extended
object, such as Fornax, one must actually compare ``center-of-mass''
(COM) motions to avoid any projection effects and internal galaxy
motions (e.g., galactic rotation) that might alter the observed
motions. For computing galactic orbits of these galaxies we also need
COM velocities. Our measurements are performed, however, on fields
away from the COM, and we have to apply corrections to account for
this situation. As explained in \citet{cos09}, in the case of the LMC
our PMs had to be corrected not only for projection effects (see
below), but also by the effect on our measured PMs induced by the
(differential) rotation of the plane of the LMC. To apply these
corrections, we followed the prescriptions described by \cite{jon94}
and \cite{van02}, which assumed that our fields lay in the plane of
the LMC (or SMC) and shared the motion of its disk. The situation for
the dSphs is quite different however, they do not exhibit any hint of
large-scale rotation, nor of the existence of a disk or other
well-defined structure; rather they show a smooth spheroidal
distribution and the motion of its stars is dominated by their
velocity dispersion, which is typically a few~km~s$^{-1}$
\citep{van99}. This velocity dispersion translates into a PM
dispersion of a few $\mu$as~y$^{-1}$ which is not measurable by
current astrometric techniques (although it will be measured in the
future by SIM). Therefore, if we assume that there are no large-scale
streaming motions in Fornax, our PMs are not affected, as far as we
can measure it, by internal kinematic effects. We do, however, need to
correct for purely geometrical projection effects, this is done as
described in the following paragraph.

If $(v_r,v_\alpha,v_\delta)$ are the observed (measured) Heliocentric
radial velocity, velocity in RA and velocity in DEC respectively, for
a field at position $(\alpha,\delta)$ and Heliocentric distance $r$,
the velocity for the COM $(v^*_r,v^*_\alpha,v^*_\delta)$ at position
$(\alpha^*,\delta^*)$ and distance $r^*$ is given by:

\begin{equation}
\left( \begin{array}{c}
v^*_r \\
v^*_\alpha \\
v^*_\delta
\end{array} \right)
= M^{-1} (\alpha^*,\delta^*) \cdot M(\alpha,\delta) \cdot
\left( \begin{array}{c}
v_r \\
v_\alpha \\
v_\delta
\end{array} \right) \label{proj}
\end{equation}

where $M(\alpha,\delta)$ is a rotation matrix whose components are:

\begin{equation}
M(\alpha,\delta) =
\left( \begin{array}{ccc}
\cos \delta \cos \alpha & -\sin \alpha  & -\sin \delta \cos \alpha \\
\cos \delta \sin \alpha &   \cos \alpha & -\sin \delta \sin \alpha \\
\sin \delta             & 0              &  \cos \delta
\end{array} \right) \label{matrix}
\end{equation}

and which satisfies that
$M^{-1}(\alpha,\delta)=M^t(\alpha,\delta)$. While $r$ and $r^*$ are
not explicitely written in equation~(\ref{proj}), they are implicitly
used to go from PMs to tangential velocities (or vice versa) through
equation~(\ref{vel}). Of course, $r^* = 138 \pm 8$~kpc. In the case of
the LMC (and SMC) \citet{cos09} assumed that our fields were located
in a disk-like structure with known orientation in the sky
(inclination and line of nodes). We do not have however such
structures in the featureless dSphs. If we assume, e.g.,
as a first crude approximation that the galaxy lies in a plane in the sky
perpendicular to the line-of-sight to the COM. In this case one can
show that $r$ is given by:

\begin{equation}
r = \frac{r^*}{\cos \delta \cos \alpha \cos \delta^* \cos \alpha^* +
  \cos \delta \sin \alpha \cos \delta^* \sin \alpha^* + \sin \delta
  \sin \delta^*}
\end{equation}

We note that, in equation~(\ref{proj}) $v_r$ is {\it not} known, but
we do know $v^*_r = 53 \pm 3$~km~s$^{-1}$ from the
literature. Therefore, starting from our measured $v_\alpha$ and
$v_\delta$ values we iterate on the $v_r$ values until we reproduce
the expected $v^*_r$ for the COM, in a procedure similar to that
adopted by \citet{cos09}. Also, equations (\ref{proj}) and
(\ref{matrix}) easily allow us to fully propagate errors on all the
measured quantities (radial velocities, proper motions, distances)
from the observed values to the sought-for COM values. For our
measured PM, the COM distance and radial velocity indicated
previously, and the Fornax COM position given by \citet{mat98}, we
obtain $v^*_{\alpha}=419 \pm 52$~km~s$^{-1}$ and $v^*_{\delta}= -7 \pm
72$~km~s$^{-1}$, i.e., a value similar to that computed without {\it
  any} projection correction. This is due to the fact that, in this
particular case, these corrections are actually quite small, the PM
corrections are 0.05~$\mu$as in RA and by 0.45~$\mu$as in DEC. These
corrections are, of course, much smaller than the measurement
uncertainties involved (this was not case of the LMC and SMC fields
reported in \citet{cos09}). Note that the field reported here is quite
close to the Fornax COM, for some of the other more distant QSO fields
in our program, these corrections might be slightly larger, but can be
readily computed in each case. None of the values reported by other
authors in Table~\ref{pmtab} have been corrected for any projection
effect, but since these corrections are tiny, one can readily compare
them without further corrections.

For completeness, in Table~\ref{pmtab} (and in Figure~\ref{pmcomp}),
we have included the recent determination of the Fornax PM by
\citet{wal08}, using what they call the ``perspective rotation''
method (in a way, this method is the reverse of the ``astrometric
radial velocities'', described by, e.g., \citet{lin00}). As it can be
easily seen from Equations~\ref{proj} and~\ref{matrix}, one could
write an equivalent equation for $(v_r, v_\alpha, v_\delta)$ as a
function of $(v^*_r, v^*_\alpha, v^*_\delta)$ by a simple matrix
inversion. The first row of that equation would give the observed
radial velocity $v_r$ as a function of a combination of $(v^*_r,
v^*_\alpha, v^*_\delta)$ (which, for a given galaxy is of course a
fixed quantity - independent of the field observed). Therefore, if one
measures the $v_r$ of samples of stars at different locations
$(\alpha,\delta)$ across a galaxy, one can solve through some
minimization algorithm for the unknown $(v^*_r, v^*_\alpha,
v^*_\delta)$. \citet{wal08} have used precisely this approach to
determine the ``perspective'' PMs for 4 dSphs, including Fornax, using
radial velocities exclusively. As it can be seen from
Table~\ref{pmtab}, the method has errors compatible with the best
purely astrometric determinations, and can thus become a useful
complement to them.

\newpage

\acknowledgments

RAM and EC acknowledge support by the Fondo Nacional de
Investigaci\'on Cient\'{\i}fica y Tecnol\'ogica (Fondecyt project
No. 1070312), the Chilean Centro de Astrof\'{\i}sica (FONDAP project
No. 15010003) and the Chilean Centro de Excelencia en Astrof\'{\i}sica
y Tecnolog\'{\i}as Afines (PFB 06). MHP acknowledges support by
Project \# 4721-09 from Universidad de Tarapac\'a. CG acknowledges
support by the Instituto de Astrof\'isica de Canarias (P3-94) and by
the Ministry of Education and Research of the Kingdom of Spain
(AYA2004-06343). RAM acknowledges extensive discussions on the early
phases of this project with the late Prof. Claudio Anguita; we are all
sorry that he never saw his ideas realized in practice, and with
graduate student Matias Jones. We are greatful to the ESO OPC for
their continued support of this long-term program, as well as to the
La Silla Scientists, Engineers and Operations staff for their
continuous help in the course of the program, specially Dr. Michael
F. Sterzik and Mr. Federico Fox. The authors would also like to thank
an anonymous refereee for his/her very helpful comments.

{\it Facilities:} \facility{ESO-NTT (SuSI2)}.


\newpage

\begin{table}
\begin{center}
\caption{Observational material used for the Fornax field centered on QSO J0240-3434B.\label{frames}}
\begin{tabular}{cccccc}
\tableline\tableline
Obs. date        & Filter & Exp. time & Hour Angle    & FWHM & Usage \\
yyyy-mm-dd  &        & s        & hh:mm & arcsec  &    \\
\tableline
2000-08-08  & B\#811 & 600      & -3:40 & 0.9 & CMD    \\
\tableline
2000-08-08  & R\#813 & 360      & -1:31 & 0.6 & Astrom \\
2000-08-08  & R\#813 & 360      & -1:22 & 0.5 & Astrom \\
2000-08-08  & R\#813 & 250      & -0:18 & 0.5 & Astrom, CTB-1 \\
2000-08-08  & R\#813 & 250      & -0:13 & 0.5 & Astrom, CTB-2 \\
2000-08-08  & R\#813 & 250      & -0:05 & 0.5 & Astrom \\
2000-08-08  & R\#813 & 300      &  0:05 & 0.4 & Astrom \\
2000-08-08  & R\#813 & 400      &  0:15 & 0.4 & Astrom \\
\tableline
2004-10-05  & R\#813 & 900      &  0:45 & 0.6 & Astrom, SFR, Master \\
2004-10-05  & R\#813 & 900      &  1:01 & 0.6 & Astrom, SFR      \\
2004-10-05  & R\#813 & 900      &  1:17 & 0.6 & Astrom, SFR      \\
\tableline
2005-01-13  & R\#813 & 900      &  1:11 & 0.8 & Astrom      \\
2005-01-13  & R\#813 & 900      &  1:27 & 0.7 & Astrom      \\
\tableline
2007-11-07  & R\#813 & 900      & -4:01 & 0.7 & DCR         \\
2007-11-07  & R\#813 & 900      & -3:40 & 0.7 & DCR         \\
2007-11-07  & R\#813 & 900      & -3:23 & 0.7 & DCR         \\
2007-11-07  & R\#813 & 900      & -3:07 & 0.7 & DCR         \\
2007-11-07  & R\#813 & 900      & -2:51 & 0.6 & DCR         \\
2007-11-07  & R\#813 & 900      & -2:34 & 0.6 & DCR         \\
2007-11-07  & R\#813 & 900      & -2:18 & 0.6 & DCR         \\
2007-11-07  & R\#813 & 900      & -2:01 & 0.8 & DCR         \\
2007-11-07  & R\#813 & 900      & -1:45 & 0.8 & DCR, Astrom        \\
2007-11-07  & R\#813 & 900      & -1:28 & 0.8 & DCR, Astrom         \\
2007-11-07  & R\#813 & 900      & -1:12 & 0.8 & DCR, Astrom         \\
2007-11-07  & R\#813 & 900      & -0:55 & 0.7 & DCR, Astrom         \\
2007-11-07  & R\#813 & 900      & -0:38 & 0.6 & DCR, Astrom         \\
\tableline
\end{tabular}
\end{center}
\end{table}

\clearpage

\begin{table}
\begin{center}
\caption{Proper motion for QSO J0240-3434B under different circumstances explained in detail in the text.\label{pms}}
\begin{tabular}{ccc}
\tableline\tableline
\mua    & \mud    & Comments \\
$\masy$   & $\masy$   &           \\
\tableline
$-0.64 \pm 0.08$ & $0.05 \pm 0.11$ & All 260 (initial) LRS stars \\
$-0.66 \pm 0.08$ & $0.04 \pm 0.11$ & 14 LRS stars with $\mu> 2.0$~$\masy$ eliminated (246 LRS stars) \\
$-0.67 \pm 0.08$ & $0.01 \pm 0.11$ & 7 stars with high registration residuals eliminated (239 LRS stars) \\
$-0.64 \pm 0.08$ & $0.01 \pm 0.11$ & 22 stars eliminated based on CMD \& high PM errors (217 LRS stars) \\
\tableline
$-0.63 \pm 0.07$ & $0.05  \pm 0.10$ & $|HA| \le 2.0$~hr (17 frames) \\
$-0.71 \pm 0.06$ & $0.08  \pm 0.09$ & $|HA| \le 3.0$~hr (21 frames) \\
$-0.71 \pm 0.17$ & $-0.01 \pm 0.11$ & $|HA| \le 1.5$~hr, no DCR correction \\
\tableline
$-0.84 \pm 0.08$ & $0.13 \pm 0.10$  & 2$^{nd}$ order registration \\ 
$-0.57 \pm 0.07$ & $-0.10 \pm 0.11$ & 4$^{th}$ order registration \\
$-0.63 \pm 0.08$ & $0.03 \pm 0.11$  & 3$^{rd}$ order registration, no $Y^3$ term in X, no $X^3$,$XY^2$ terms in Y \\
$-0.58 \pm 0.08$ & $-0.10 \pm 0.14$ & Delete one every-other LRS star (109 LRS stars) \\
\tableline
\end{tabular}
\end{center}
\end{table}

\clearpage

\begin{table}
\begin{center}
\caption{Proper motion for Fornax from different authors.\label{pmtab}\newline}
\begin{tabular}{ccc}
\tableline\tableline
\mua    & \mud    & Reference \\
$\masy$   & $\masy$   &           \\
\tableline
$0.64 \pm 0.08$ & $-0.01 \pm 0.11$ & This work (SuSI2) - J0240-3434B \\
\tableline
$0.28 \pm 0.30$ & $-0.45 \pm 0.28$ & Dinescu (plates+HST) - mean of 8 QSOs \\
$0.70 \pm 0.18$ & $-0.01 \pm 0.19$ & Dinescu (plates+HST) - mean of 48 Galaxies \\
\tableline
$0.541 \pm 0.085$ &  $-0.275 \pm 0.071$ & Piatek (HST) - J0240-3434A \\
$0.424 \pm 0.096$ &  $-0.477 \pm 0.109$ & Piatek (HST) - J0240-3434B \\
$0.536 \pm 0.158$ &  $-0.325 \pm 0.162$ & Piatek (HST) - J0240-3438 \\
$0.446 \pm 0.072$ &  $-0.391 \pm 0.060$ & Piatek (HST) - J0238-3443 \\
\tableline
$0.48 \pm 0.15$ &  $-0.25 \pm 0.14$ & Walker - ``perspective'' motions \\
\tableline
\end{tabular}
\end{center}
\end{table}

\clearpage

\clearpage

\clearpage

\clearpage

\clearpage

\clearpage

\clearpage

\clearpage

\clearpage

\begin{figure}
\epsscale{.80}
\plotone{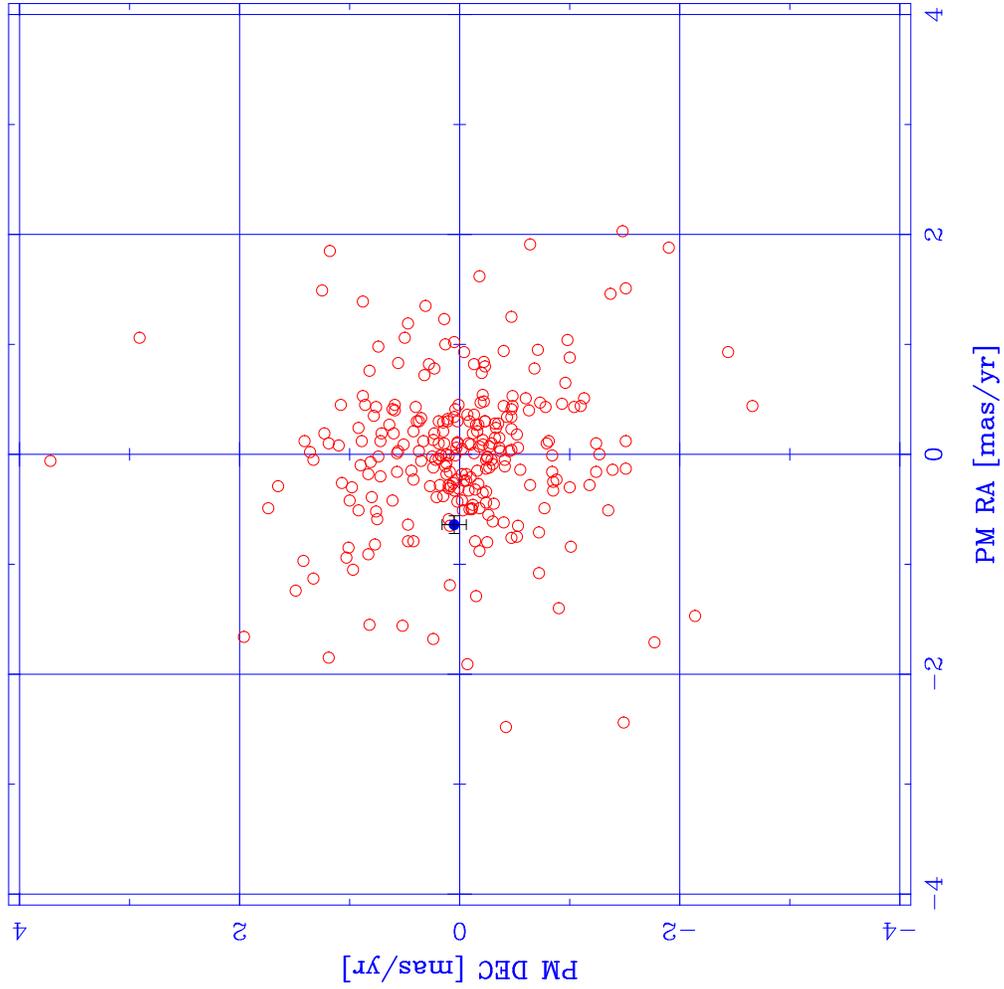}
\caption{Vector point diagram based on all the initial 260 LRS
  stars. The dot with error bars represent the QSO. \label{pmall}}
\end{figure}

\clearpage

\begin{figure}
\epsscale{.80}
\plotone{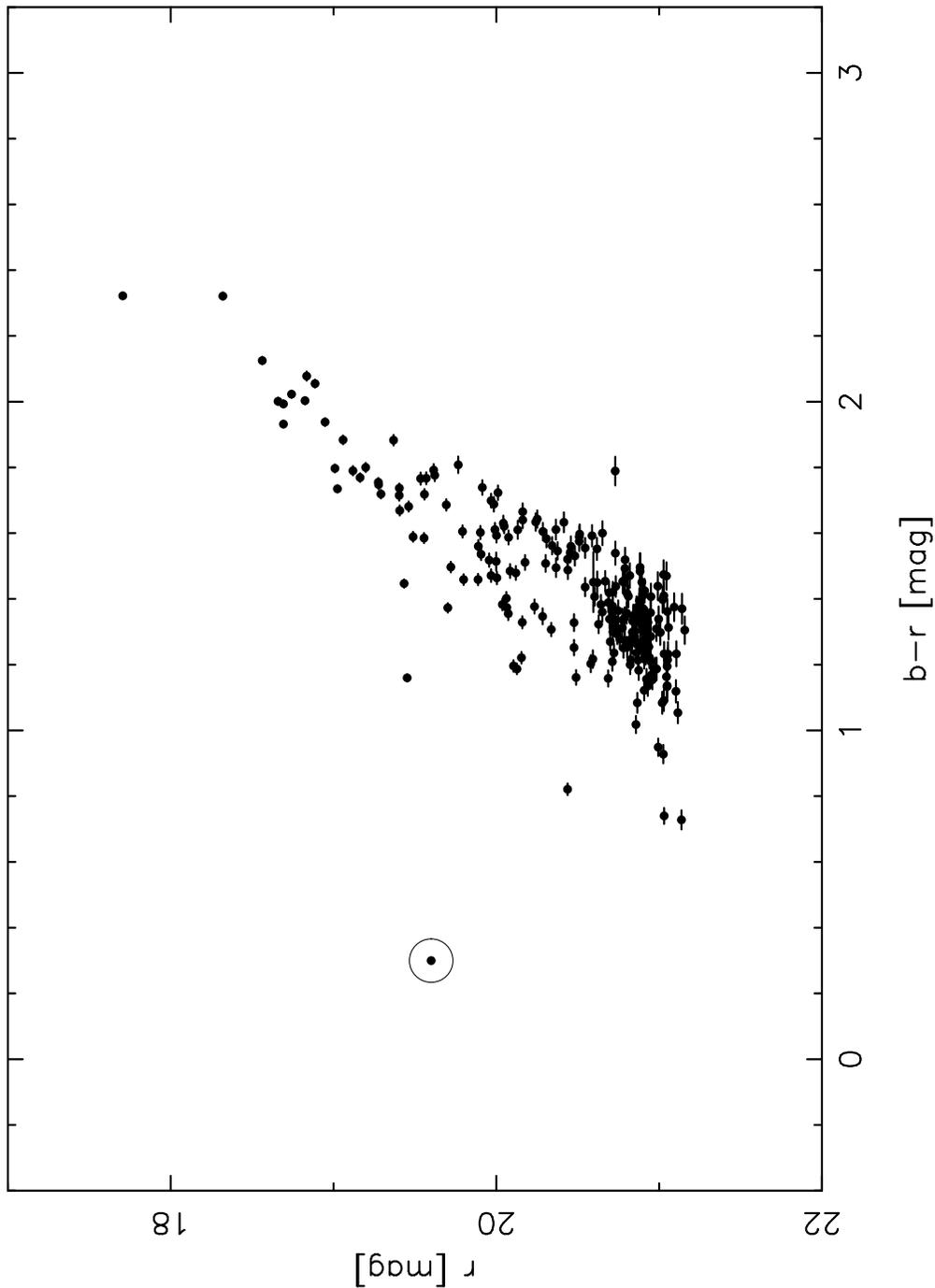}
\caption{CMD for our clean sample (free of high PM stars and
  registration outliers) comprising 239 LRS stars (dots with PSF
  photometric error bars), and QSO~J0240-3434B (circled dot). This
  figure, when compared, e.g., with the Fornax deep photometry in
  Figure~7 by \citet{ste97} indicates that Galactic contamination is
  minimal. This figure allows us to eliminate LRS objects that due to
  their magnitude-color combination are unlikely members of
  Fornax. Moreover, if we compare this figure with, e.g., the Fornax
  deep photometry in Figure~7 by \citet{ste97} we can see that
  Galactic contamination is minimal for this field, in agreement with
  our estimates based on the starcount Galactic model by \citet{men96}
  (see text). \label{cmd1} }
\end{figure}

\clearpage

\begin{figure}
\epsscale{.80}
\plotone{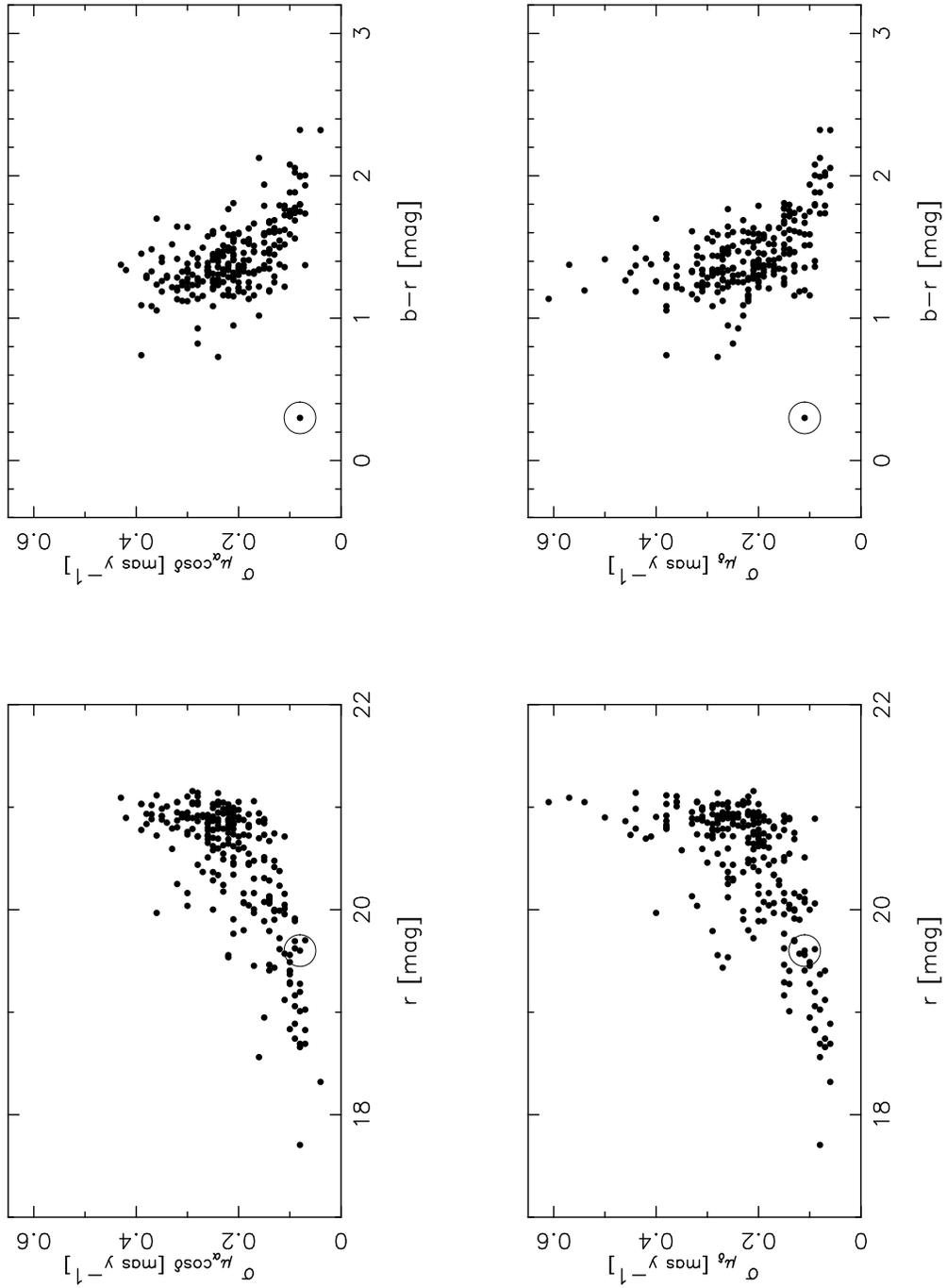}
\caption{PM errors for the same sample of LRS as in Figure~\ref{cmd1}
  as a function of magnitude and color. The circled-dot is the
  QSO. This figure allows us to eliminate LRS objects with unusually
  large PM errors for their magnitude. \label{errpm1}}
\end{figure}

\clearpage

\begin{figure}
\epsscale{.80}
\plotone{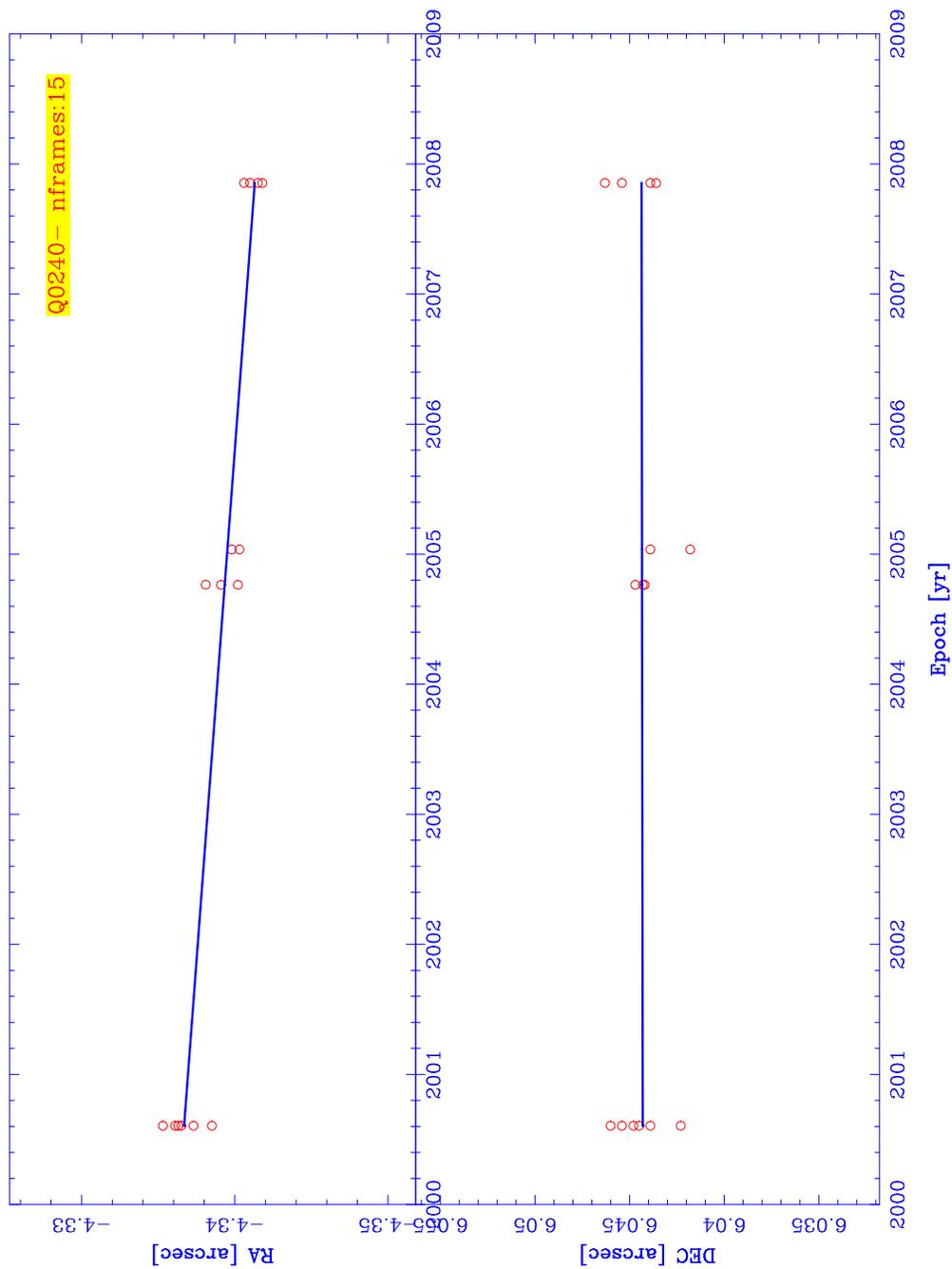}
\caption{Barycentric position (DCR-corrected) {\it vs.}  epoch diagram
  for QSO~J0240-3434B. The slope is the PM of the QSO with respect to
  the LRS stars described in the text. The 15 frames indicated in
  Table~\ref{frames} with $|HA|< 1.5$~hrs went into this solution. The
  overall rms of the position {\it vs.} epoch fits are (0.88,1.27)~mas
  in RA \& DEC respectively. \label{qsopm}}
\end{figure}

\clearpage

\begin{figure}
\epsscale{.80}
\plotone{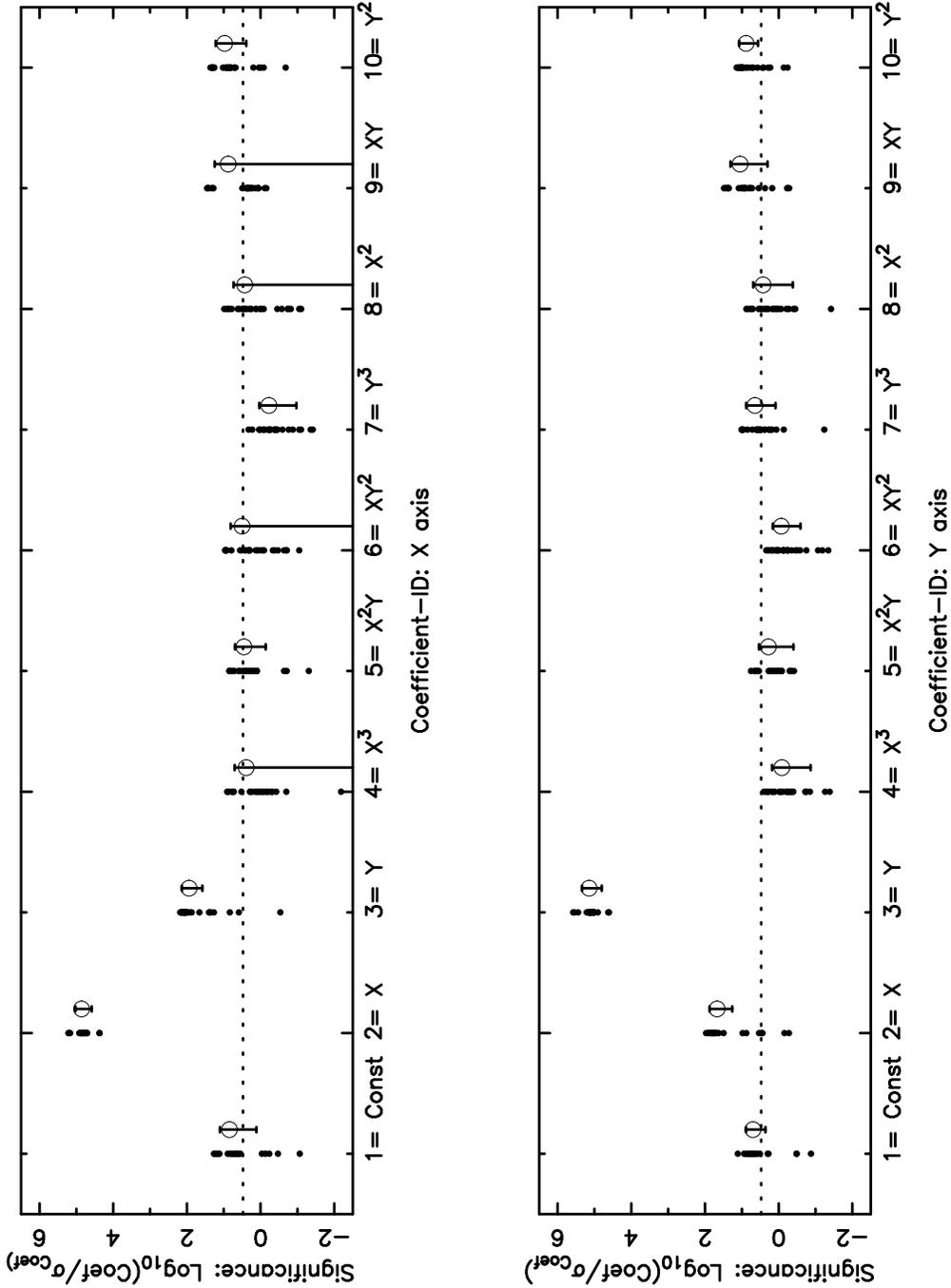}
\caption{``Significance'' (see text for its definition) of each of the
  polynomial terms (indicated on the abscissa) for our 3$^{rd}$ order
  registration. Each point represents one of the 15 frames registered
  into the SFR. On the right side of the distribution of points we
  give, for each coefficient, the mean of the significance and its
  1$\sigma$ error interval (the error bar). The dotted line represents
  a significance, in this logarithmic scale, of $\sim0.477$,
  equivalent to a coefficient that is 3$\times$ larger than its
  (formal) error from the $\chi^2$ fit. When the average significance
  (and its interval) are below the dotted line, this indicates
  non-significant coefficient.\label{coef}}
\end{figure}

\clearpage

\begin{figure}
\epsscale{.80}
\plotone{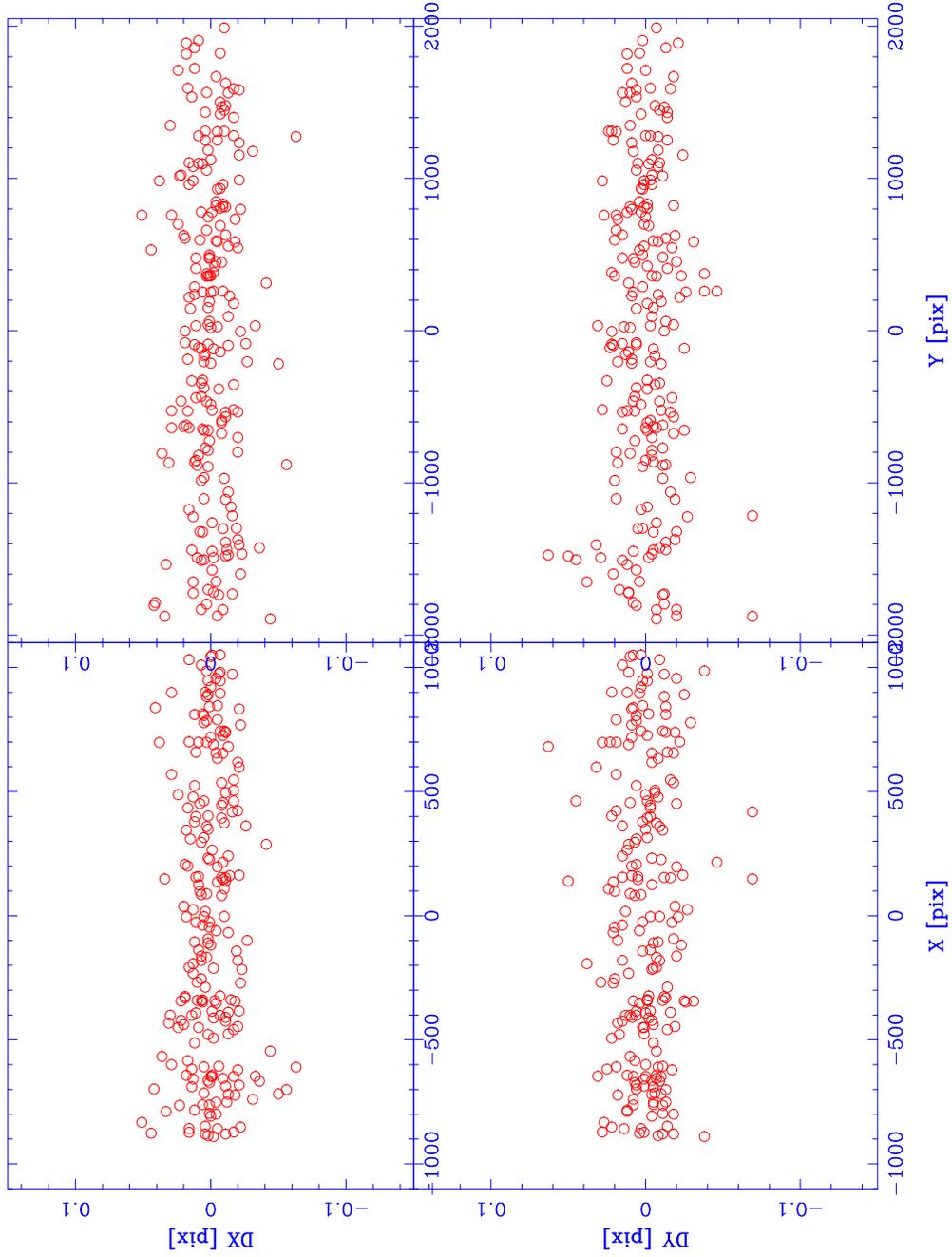}
\caption{Registration residuals for one of our frames using our
  3$^{rd}$ order registration polynomial. There are no obvious
  positional trends.\label{resi}}
\end{figure}

\clearpage

\begin{figure}
\includegraphics[angle=-90,scale=1.0]{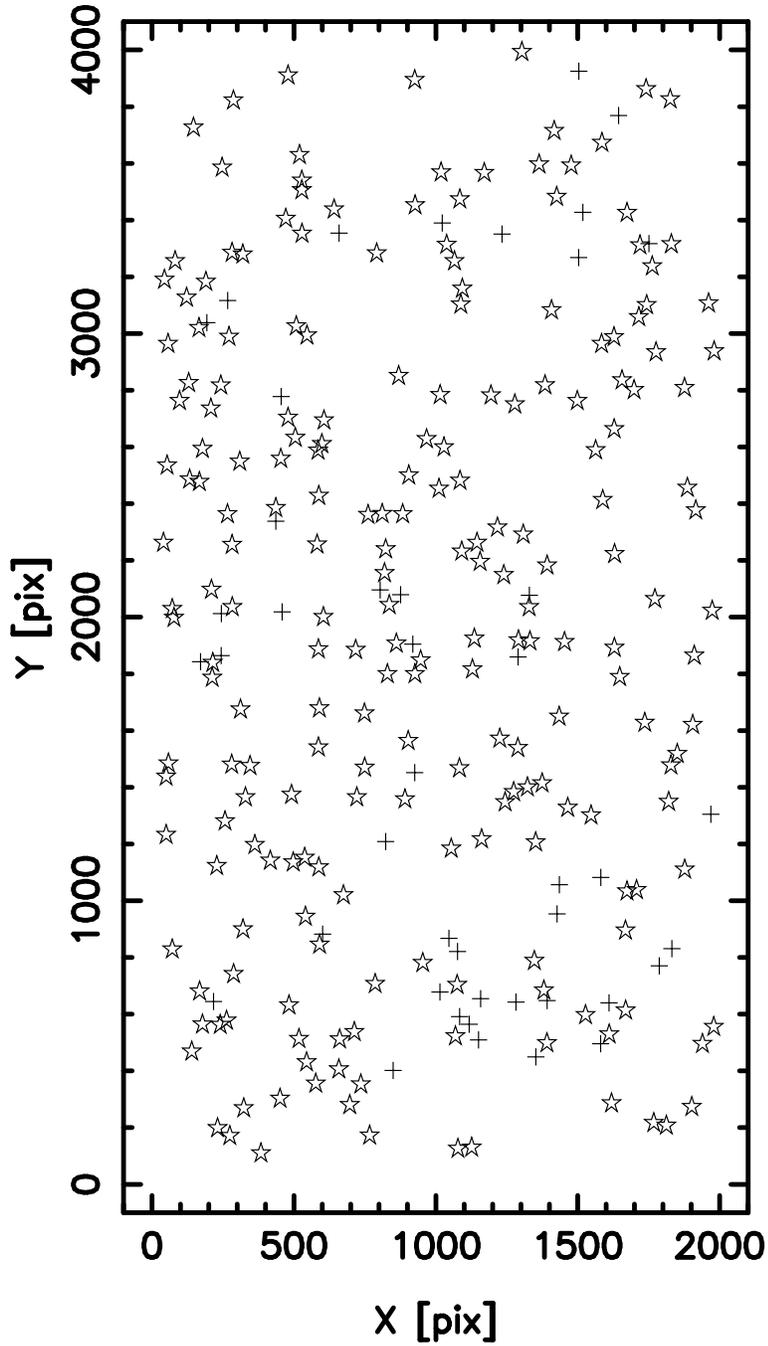}
\caption{X-Y distribution of our final 217 LRS stars (open star
  symbol). The $+$ symbol indicates 43 initial LRS stars that were
  subsequently purged for various reasons, discussed in the text. Both
  sets of stars are distributed over the entire FOV. \label{xylrs} }
\end{figure}

\clearpage

\begin{figure}
\epsscale{.80}
\plotone{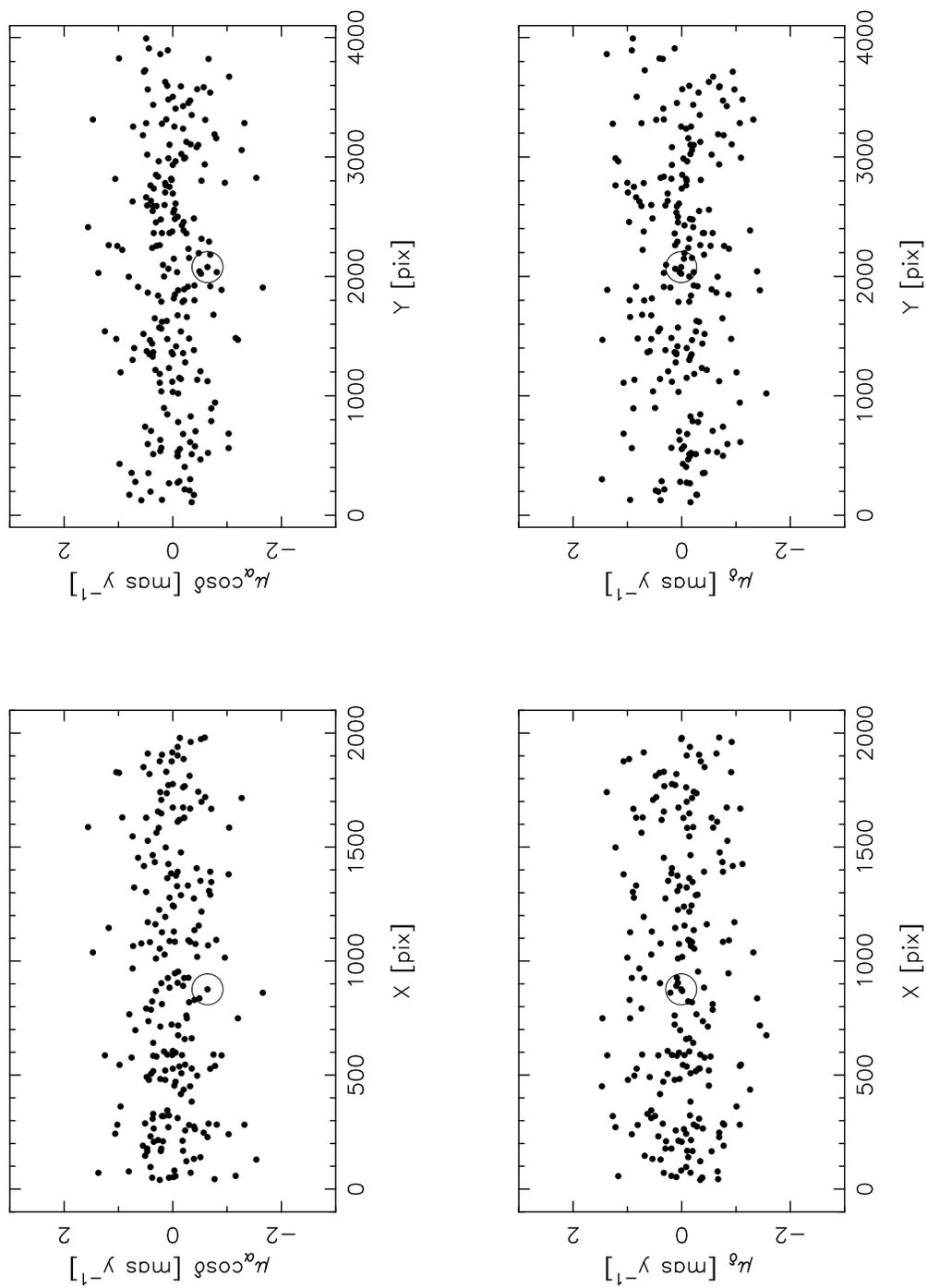}
\caption{Proper motions {\it vs.} X- and Y-coordinates for the final
  217 LRS stars (dots) and the QSO (circle). No obvious trends {\it
    vs.}  position in the frames are discernible. \label{pmxy} }
\end{figure}

\clearpage

\begin{figure}
\epsscale{.80}
\plotone{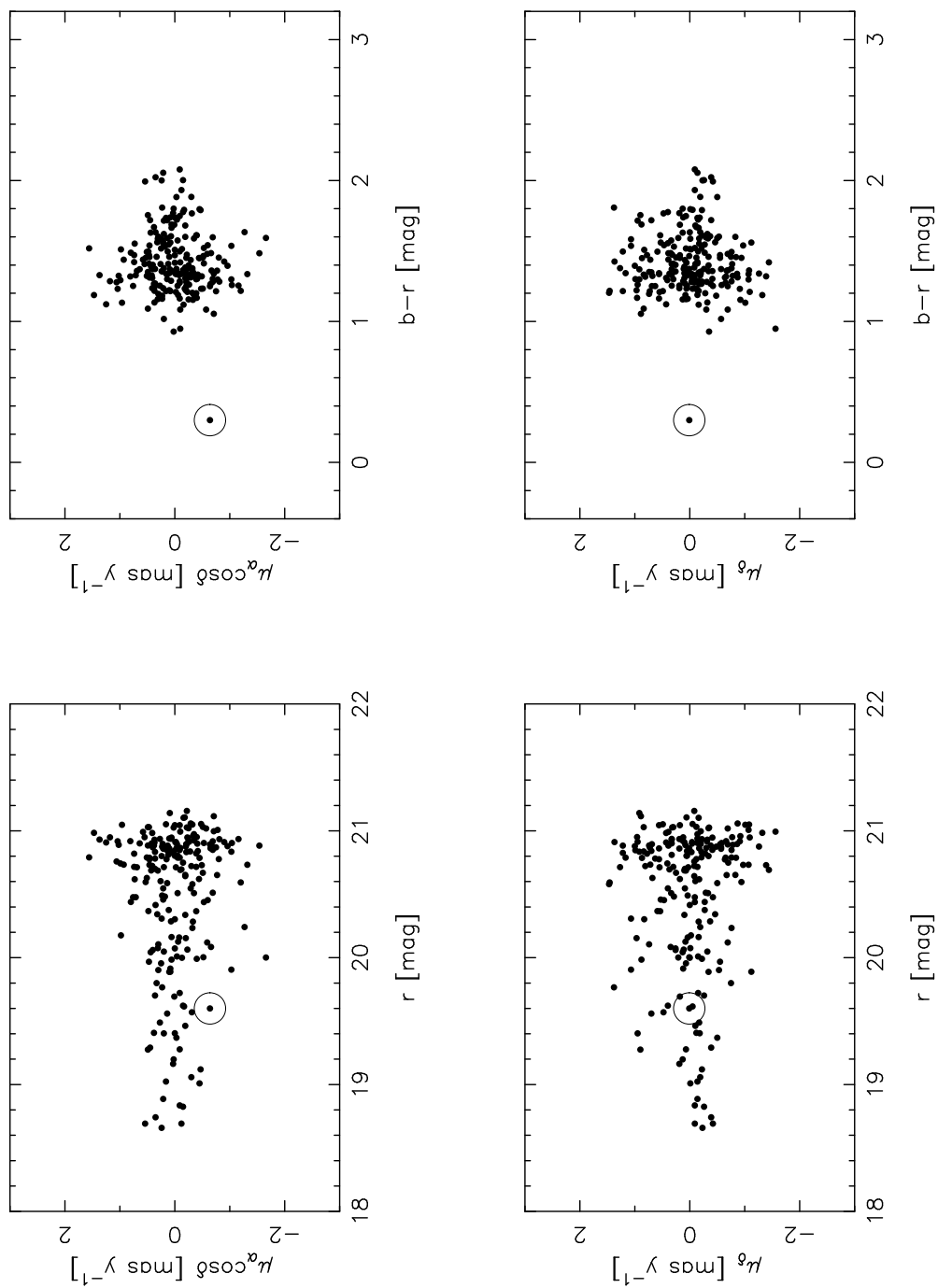}
\caption{Same as Figure~\ref{pmxy} but {\it vs.} magnitude and
  color. Other than the expected increase in (pseudo-) PMs for the LRS
  stars (see text) at fainter magnitudes, and the much bluer color of
  the QSO with respect to the Fornax stars, no obvious trends are
  found. \label{pmmag} }
\end{figure}

\clearpage

\begin{figure}
\epsscale{.80}
\plotone{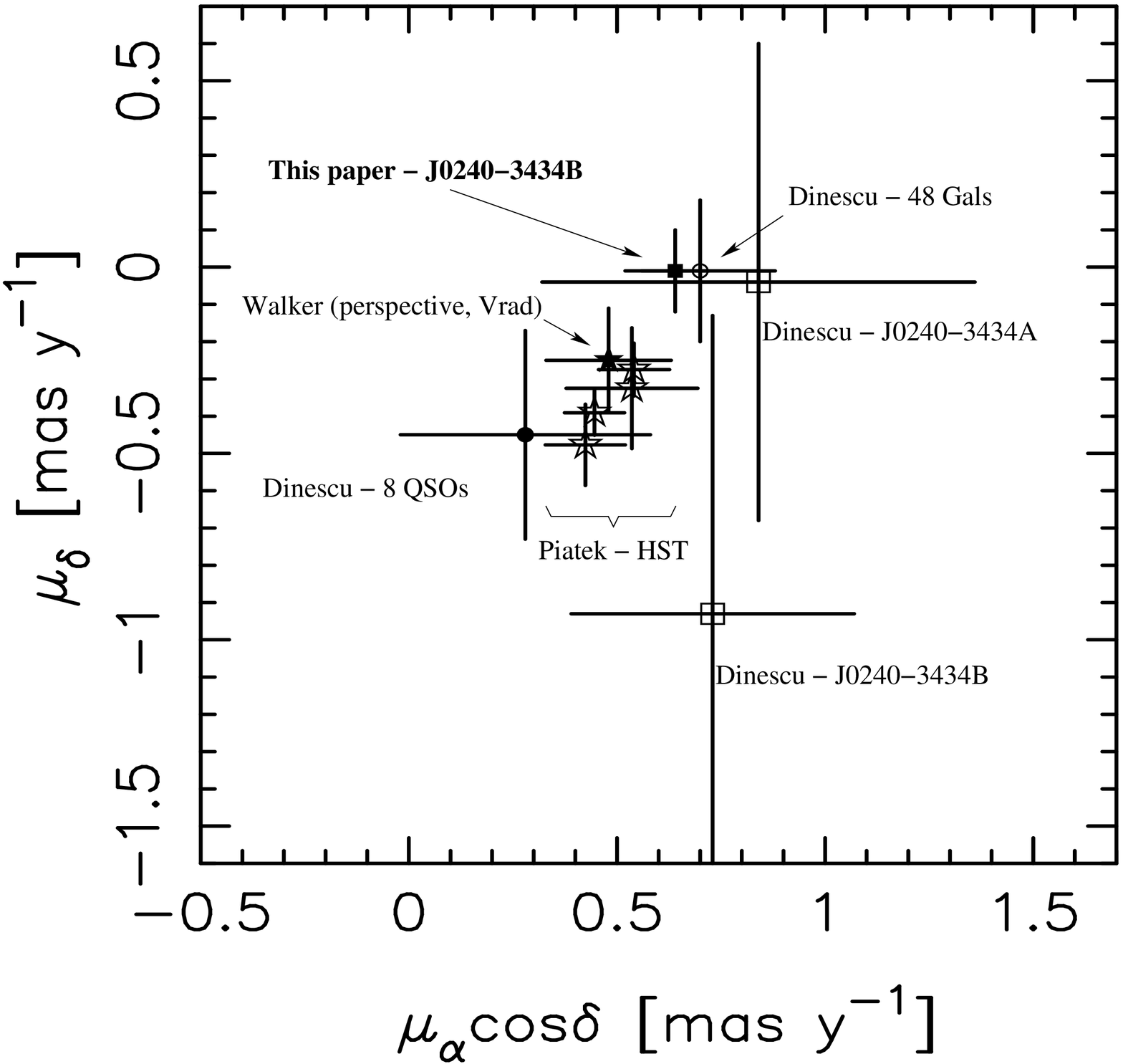}
\caption{Comparison of Fornax PMs. The open stars are the HST values
  from \citet{pia07} from 4 individual QSOs, the filled and open
  circle are the mean PMs using 8 QSOs and 48 galaxies respectively
  from the Plates+HST/WFPC study from \citet{din04}, and the filled
  square is our single measurement on QSO~J0240-3434B. The open
  squares are the individual measurements reported in the
  \citet{din04} paper for both components (image A and B) of the same
  QSO reported by us in this study. We also include (filled star) the
  recent result by \citet{wal08} (see Table~\ref{pmtab}) obtained
  through the ``perspective'' proper motion method using radial
  velocities alone, and further discussed in the text (note that, in a
  way, this should be considered a measurement based on a weighted
  mean of many lines of sight - using in total 2,610 Fornax stars). As
  it can be seen our value is closer to the \citet{din04}
  determination using {\it galaxies} rather than their value using
  QSOs. Also, our single measurement errors are similar to those from
  HST. \label{pmcomp} }
\end{figure}

\end{document}